%% file: main.tex
\RequirePackage{fix-cm}
\documentclass{svjour3}                     
\smartqed  
\usepackage{graphicx}

\usepackage[ruled,vlined,linesnumbered]{algorithm2e}

\usepackage[bookmarks,bookmarksopen,bookmarksdepth=2]{hyperref}

\usepackage{filecontents}

\input{data-num-time.tex}
\input{data-trunk-size.tex}
\input{data-stability.tex}
\input{data-sample.tex}
\input{figdef.tex}

\usepackage{soul,color}

\newcommand{\del}[1]{{}}

\usepackage{mathrsfs}
\usepackage{pifont}
\newcommand{\cmark}{\ding{51}}%
\newcommand{\xmark}{\ding{55}}%

\usepackage{todonotes}

\usepackage{amssymb}
\usepackage{autobreak}
\usepackage{breqn}

\usepackage{setspace}
\usepackage{enumitem}

\usepackage{hyperref}

\newcommand{\toprule}{\hline\noalign{\smallskip}}
\newcommand{\midrule}{\noalign{\smallskip}\hline\noalign{\smallskip}}
\newcommand{\bottomrule}{\noalign{\smallskip}\hline}
\usepackage{multirow}

\usepackage{pgfplots}
\usepackage{pgfplotstable}
\usepackage{csvsimple} 

\usepackage{pgfkeys}
    \newenvironment{customlegend}[1][]{%
        \begingroup
        \csname pgfplots@init@cleared@structures\endcsname
        \pgfplotsset{#1}%
    }{%
        \csname pgfplots@createlegend\endcsname
        \endgroup
    }%
    \def\addlegendimage{\csname pgfplots@addlegendimage\endcsname}

\usepackage{textcomp}

\newcommand{\DEAD}{{\texttt{DEAD}}}
\newcommand{\LIVE}{{\texttt{LIVE}}}
\newcommand{\TRUE}{{\texttt{TRUE}}}
\newcommand{\FALSE}{{\texttt{FALSE}}}

\newcommand{\status}{{\it status}}
\newcommand{\degout}{{\it deg_{out}}}
\newcommand{\degin}{{\it deg_{in}}}
\newcommand{\Degout}{{\it Deg_{out}}}
\newcommand{\Degin}{{\it Deg_{in}}}
\newcommand{\post}{{\it post}}
\newcommand{\pre}{{\it pre}}
\newcommand{\revised}{{\it revised}}
\newcommand{\change}{{\it change}}
\newcommand{\lock}{{\it lock}}

\begin{document}
\pagestyle{plain}



\title{Efficient Parallel Graph Trimming by Arc-Consistency}

\author{Bin Guo         \and
        Emil Sekerinski 
}


\institute{Bin Guo \at
              Department of Computing and Software, McMaster University, Hamilton, Ontario\\
              Tel.: +1204-9529589\\
              \email{guob15@mcmaster.ca}            
           \and
           Emil Sekerinski \at
              Department of Computing and Software, McMaster University, Hamilton, Ontario \\
              \email{emil@mcmaster.ca}\\
}

\date{Received: date / Accepted: date}

\maketitle

\begin{abstract}
Given a large data graph, trimming techniques can reduce the search space by removing vertices without outgoing edges. One application is to speed up the parallel decomposition of graphs into strongly connected components (SCC decomposition), which is a fundamental step for analyzing graphs. We observe that graph trimming is essentially a kind of arc-consistency problem, and AC-3, AC-4, and AC-6 are the most relevant arc-consistency algorithms for application to graph trimming. 
The existing parallel graph trimming methods require worst-case $\mathcal O(nm)$ time and worst-case $\mathcal O(n)$ space for graphs with $n$ vertices and $m$ edges. We call these parallel AC-3-based as they are much like the AC-3 algorithm.
In this work, we propose AC-4-based and AC-6-based trimming methods. That is, AC-4-based trimming has an improved worst-case time of $\mathcal O(n+m)$ but requires worst-case space of $\mathcal O(n+m)$; compared with AC-4-based trimming, AC-6-based has the same worst-case time of $\mathcal O(n+m)$ but an improved worst-case space of $\mathcal O(n)$. 
We parallelize the AC-4-based and AC-6-based algorithms to be suitable for shared-memory multi-core machines. The algorithms are designed to minimize synchronization overhead. 
For these algorithms, we also prove the correctness and analyze time complexities with the work-depth model. 

In experiments, we compare these three parallel trimming algorithms over a variety of real and synthetic graphs on a multi-core machine, where each core corresponding to a worker. Specifically, for the maximum number of traversed edges per worker by using 16 workers, AC-3-based traverses up to 58.3 and 36.5 times more edges than AC-6-based trimming and AC-4-based trimming, respectively. 
That is, AC-6-based trimming traverses much fewer edges than other methods, which is meaningful especially for implicit graphs. In particular, for the practical running time, AC-6-based trimming achieves high speedups over graphs with a large portion of trimable vertices.
\end{abstract}


\section*{Acknowledgment} 
We acknowledge the support of the Natural Sciences and Engineering Research Council of Canada (NSERC).

\newpage
\input{body-section}

\input{body-experiment}

\section{Conclusions}
In this work, we study graph trimming algorithms for removing vertices without outgoing edges. The arc-consistency algorithms, in particular AC-3, AC-4, and AC-6, can be applied to graph trimming, leading to the so-called AC-3-based, AC-4-based, and AC-6-based trimming algorithms, respectively.
Based on that, we propose parallel AC-4-based and AC-6-based trimming algorithms that have better worst-case time complexities than AC-3-based. For these three trimming algorithms, we summarize the trend and test results bellow: 
\begin{itemize}
    \item The common existing graph trimming method is actually parallel AC-3-based, which has worst-case time complexity. Although AC-4-based and AC-6-based algorithms have similar worst-case time complexities, the AC-6-based algorithm traverses fewer edges per worker and requires less memory usage than the AC-4-based one. 
    \item For implicit graphs in which edges are generated on-the-fly, the AC6-based algorithm does not rely on the reversed graphs, unlike the AC4-based algorithm, and thus is more suitable for trimming implicit graphs.  
    \item For explicit graphs in which all edges are linearly stored in memory, our AC-6-based algorithm does not always outperform the other methods, but always traverses the least number of edges and has the best stability and scalability. 
\end{itemize}

In future work, we can apply graph trimming to Strong Connected Components (SCC) decomposition as a great percentage of \mbox{size-$1$} SCCs can be trimmed in parallel. We also can apply graph trimming to cycle directions as the trimmable vertices cannot be in cycles and can be trimmed in parallel. 
Both applications depend on Depth First Search (DFS), which is hard to parallelize. However, our trimming techniques can efficiently trim graphs in parallel if there are a large portion of trimmable vertices in graphs.
In particular, we can apply the AC-6-based algorithm to trim the model checking graphs in which edges are expensively calculated on-the-fly; fewer traversed edges will likely save the running time.
In addition, we can device the graph trimming algorithms for distribute system to address the large scale parallelism.  

\bibliographystyle{spmpsci} 
\bibliography{ref.bib}
\end{document}

%% file: figdef.tex
\usepackage{subcaption}
\usepackage{pgfplots}
\usepackage{tikz}
\usepackage{tikzscale}
\definecolor{darkgreen}{RGB}{0,104,0}
\pgfplotsset{compat=1.9}

\newcommand{\BASEPLOT}{\addplot[color=black, mark=none, thick, error bars/.cd, y dir=both, y explicit, error bar style={thick}, error mark options={rotate=90,thick}]
}

\newcommand{\TRIMPLOT}{\addplot[color=red, mark=triangle*, thick, error bars/.cd, y dir=both, y explicit, error bar style={thick}, error mark options={rotate=90,thick}]
}
\newcommand{\FASTTRIMPLOT}     {\addplot[color=blue, mark=oplus*, thick, error bars/.cd, y dir=both, y explicit, error bar style={thick}, error mark options={rotate=90,thick}]
}

\newcommand{\ACFOURPLOT}     {\addplot[color=darkgreen, mark=square*, thick, error bars/.cd, y dir=both, y explicit, error bar style={thick}, error mark options={rotate=90,thick}]
}

\newcommand{\ACFOURPLOTS}     {\addplot[color=darkgreen, mark=square, thick, error bars/.cd, y dir=both, y explicit, error bar style={thick}, error mark options={rotate=90,thick}]
}

\newcommand{\TIMESUBFIGURE}{
        \begin{subfigure}[b]{\fwide\textwidth}
        \centering
        \resizebox{1\textwidth}{!}{
        \begin{tikzpicture}
         \begin{axis}     [
            title={\textbf{\title}},
            xlabel={Number of workers},
            xmin=0.7, xmax=42,
            ymode=log, log basis y={2},
            xmode=log, log basis x={2},
            xticklabel=\pgfmathparse{2^\tick}\pgfmathprintnumber{\pgfmathresult},
            ylabel={Time used (ms)},
            y dir=normal,
            grid style=dashed,
            grid=both,
            legend columns=-1,
        ]
        \TRIMPLOT table [x=worker, y=AC3Trim, col sep=comma, y error plus expr=(\thisrowno{2}), y error minus expr=(\thisrowno{3})] {\csvfilename}; 
        \ACFOURPLOT table [x=worker, y=AC4Trim, col sep=comma, y error plus expr=(\thisrowno{5}), y error minus expr=(\thisrowno{6})] {\csvfilename}; 
        \FASTTRIMPLOT table [x=worker, y=AC6Trim, col sep=comma, y error plus expr=(\thisrowno{8}), y error minus expr=(\thisrowno{9})] {\csvfilename};
        \end{axis}
        \end{tikzpicture}
    } 
    \end{subfigure} 
}

\newcommand{\NUMSUBFIGURE}{
    \begin{subfigure}[b]{\fwide\textwidth}
        \centering
        \resizebox{1\textwidth}{!}{
        \begin{tikzpicture}
        \begin{axis}[ 
            title={\textbf{\title}},
            xlabel={Number of workers},
            xmin=0.7, xmax=42,
            \yminmax
            ymode=log, log basis y={2},
            xmode=log, log basis x={2},
            xticklabel=\pgfmathparse{2^\tick}\pgfmathprintnumber{\pgfmathresult},
            ylabel={Number of traversed edges},
            y dir=normal,
            grid style=dashed,
            grid=both,
        ]
        
         \TRIMPLOT table [x=worker, y=AC3Trim, col sep=comma, y error plus expr=(\thisrowno{2}), y error minus expr=(\thisrowno{3})] {\csvfilename}; 
        \ACFOURPLOT table [x=worker, y=AC4TrimAll, col sep=comma, y error plus expr=(\thisrowno{5}), y error minus expr=(\thisrowno{6})] {\csvfilename}; 
        \ACFOURPLOTS table [x=worker, y=AC4Trim, col sep=comma, y error plus expr=(\thisrowno{5}), y error minus expr=(\thisrowno{6})] {\csvfilename}; 
        
        \FASTTRIMPLOT table [x=worker, y=AC6Trim, col sep=comma, y error plus expr=(\thisrowno{8}), y error minus expr=(\thisrowno{9})] {\csvfilename};
        
        \BASEPLOT table [x=worker, y=M, col sep=comma]{\csvfilename};
        \end{axis}
        \end{tikzpicture}
    } 
    \end{subfigure} 
}

\newcommand{\TRUNKSUBFIGURE}{
    \begin{subfigure}[b]{\fwide\textwidth}
        \centering
        \resizebox{1\textwidth}{!}{
        \begin{tikzpicture}
        \begin{axis}[ 
            title={\textbf{\title}},
            xlabel={Chunk size},
            xmin=0.7, xmax=1248576,
            \yminmax
            ymode=log, log basis y={2},
            xmode=log, log basis x={2},
            ylabel={Time used (ms)},
            y dir=normal,
            grid style=dashed,
            grid=both,
        ]

         \TRIMPLOT table [x=step, y=AC3Trim, col sep=comma, y error plus expr=(\thisrowno{2}), y error minus expr=(\thisrowno{3})] {\csvfilename}; 
        \ACFOURPLOT table [x=step, y=AC4Trim, col sep=comma, y error plus expr=(\thisrowno{5}), y error minus expr=(\thisrowno{6})] {\csvfilename}; 
        \FASTTRIMPLOT table [x=step, y=AC6Trim, col sep=comma, y error plus expr=(\thisrowno{8}), y error minus expr=(\thisrowno{9})] {\csvfilename};
        
        \end{axis}
        \end{tikzpicture}
    } 
    \end{subfigure} 
}

\newcommand{\STABILITYFIGURE}{
    \begin{subfigure}[b]{\fwide\textwidth}
        \centering
        \resizebox{1\textwidth}{!}{
        \begin{tikzpicture}
        \begin{axis}[ 
            title={\textbf{\title}},
            xlabel={Repeat Times},
            xmin=0.7, xmax=51,
            ymode=log, log basis y={2},
            ylabel={\myylabel},
            y dir=normal,
            grid style=dashed,
            grid=both,
        ]
        \TRIMPLOT table [x=repeat, y=AC3Trim, col sep=comma] {\csvfilename}; 
        \ACFOURPLOT table [x=repeat, y=AC4Trim, col sep=comma] {\csvfilename}; 
        \FASTTRIMPLOT table [x=repeat, y=AC6Trim, col sep=comma] {\csvfilename};
        
        \end{axis}
        \end{tikzpicture}
    } 
    \end{subfigure} 
}

\newcommand{\SAMPLEFIGURENUM}{
    \begin{subfigure}[b]{\fwide\textwidth}
        \centering
        \resizebox{1\textwidth}{!}{
        \begin{tikzpicture}
        \begin{axis}[ 
            title={\textbf{\title}},
            xlabel={\myxlabel},
            xmin=7, xmax=103,
            ymode=log, log basis y={2},
            ylabel={\myylabel},
            y dir=normal,
            grid style=dashed,
            grid=both,
        ]
        \TRIMPLOT table [x=vs, y=AC3Trim, col sep=comma, y error plus expr=(\thisrowno{2}), y error minus expr=(\thisrowno{3})] {\csvfilename}; 
        \ACFOURPLOT table [x=vs, y=AC4TrimAll, col sep=comma, y error plus expr=(\thisrowno{5}), y error minus expr=(\thisrowno{6})] {\csvfilename}; 
        \FASTTRIMPLOT table [x=vs, y=AC6Trim, col sep=comma, y error plus expr=(\thisrowno{8}), y error minus expr=(\thisrowno{9})] {\csvfilename};
        
        \end{axis}
        \end{tikzpicture}
    } 
    \end{subfigure} 
}

\newcommand{\SAMPLEFIGURE}{
    \begin{subfigure}[b]{\fwide\textwidth}
        \centering
        \resizebox{1\textwidth}{!}{
        \begin{tikzpicture}
        \begin{axis}[ 
            title={\textbf{\title}},
            xlabel={\myxlabel},
            xmin=7, xmax=103,
            ymode=log, log basis y={2},
            ylabel={\myylabel},
            y dir=normal,
            grid style=dashed,
            grid=both,
        ]
        \TRIMPLOT table [x=vs, y=AC3Trim, col sep=comma, y error plus expr=(\thisrowno{2}), y error minus expr=(\thisrowno{3})] {\csvfilename}; 
        \ACFOURPLOT table [x=vs, y=AC4Trim, col sep=comma, y error plus expr=(\thisrowno{5}), y error minus expr=(\thisrowno{6})] {\csvfilename}; 
        \FASTTRIMPLOT table [x=vs, y=AC6Trim, col sep=comma, y error plus expr=(\thisrowno{8}), y error minus expr=(\thisrowno{9})] {\csvfilename};
        
        \end{axis}
        \end{tikzpicture}
    } 
    \end{subfigure} 
}

\newcommand{\DELETENUMBAR}{
    \begin{subfigure}[b]{\fwide\textwidth}
        \centering
        \resizebox{1\textwidth}{!}{
        \begin{tikzpicture}
        \begin{axis}[ 
            title={\textbf{\title}},
            xlabel={\myxlabel},
            xmin=10, xmax=100,
            ymin=0, ymax=100,
            log ticks with fixed point,
            ylabel={Delete vertices ratio (\%)},
            xlabel={\myxlabel},
            y dir=normal,
            grid style=dashed,
            grid=both,
        ]

        \addplot[color=red, mark=triangle*,solid] table [x=vs, y=f, col sep=comma] {\csvfilename}; 
        \addplot[color=darkgreen, mark=square*,solid] table [x=vs, y=t, col sep=comma] {\csvfilename}; 
        \addplot[color=blue, mark=oplus*, solid] table [x=vs, y=tm, col sep=comma] {\csvfilename};
        
        \end{axis}
        \end{tikzpicture}
    } 
    \end{subfigure} 
}

%% file: body-section.tex
\section{Introduction}
In numerous applications, like social networks~\cite{takac2012data}, pattern matching~\cite{chen2019delta}, communication networks~\cite{kumar2010structure}, knowledge graphs~\cite{xiaoping2021construction}, and model verification~\cite{hojati1993bdd}, data is organized into directed graphs with vertices for objects and edges for their relationships.
The large size of such graphs motivates graph \emph{trimming}, i.e. removing vertices without outgoing edges to speed up subsequent processing, such as cycle detection \cite{Lowe2016}, $k$-core decomposition \cite{bz2003}, and in particular~graph decomposition~\cite{hong2013fast}. For instance, for the communication network \emph{wiki-talk}~\cite{kumar2010structure} with 2.4 million vertices, surprisingly 94.5\% of the vertices can be trimmed, which greatly reduces the graph size for subsequent processing.

One issue is that trimming such unqualified vertices may cause other vertices to become useless. Naively repeating the trimming process may lead to a quadratic worst-case time complexity. Thus, linear time bounded graph trimming methods are desired. Additionally, the availability of multi-core processors motivates efficient parallelization of such graph trimming methods. Here, a \emph{worker} is a working process corresponding to a physical core for a multi-core processor.   

To the best of our knowledge, there exists little work on parallel trimming over large data graphs, except for~\cite{McLendon2005,hong2013fast,Slota2014,iSpanJi,Chen2018}. In these studies, the graph trimming is adopted to quickly remove the vertices without out-going edges so that can speed up the strongly connected component (SCC) decomposition.
In~\cite{McLendon2005}, McLendon et al.~first apply a linear time graph trimming method to remove \emph{\mbox{size-$1$}} SCCs; however, a parallel version is not provided.
In~\cite{hong2013fast}, Hong et al.~propose a quadratic time graph trimming technique by ``peeling'' \emph{\mbox{size-$1$}} and \emph{\mbox{size-$2$}} SCCs, i.e.~SCCs with only 1 or 2 vertices. 
The ``peeling'' step is straightforward: (1)~all vertices are checked in parallel and the \emph{trimmable} ones are removed, which may cause other vertices to become trimmable; (2)~this process is repeated until no vertex can be removed from the graph. 
The advantage of this graph trimming technique is that it can be highly parallelized without difficulties.
However, it has a quadratic worst-case time complexity of $\mathcal{O}(n m / \mathcal{P}+\alpha)$, where $n$ is the number of vertices, $m$ is the number of edges, $\mathcal{P}$ is the number of workers, and $\alpha$ is the depth of the algorithm (explained in the next section). 
This parallel trimming technique is widely used in later SCC decomposition methods~\cite{Slota2014,iSpanJi,Chen2018}. 

In this work, we apply the well-known \emph{arc-consistency} (AC) algorithms to graph trimming. Based on that, we not only classify existing graph trimming algorithms but also propose a new graph trimming algorithm that improves the time and space complexities by an order of magnitude.
Before discussing these contributions, we first show an application of graph trimming, the SCC decomposition in large graphs~\cite{McLendon2005,hong2013fast,Slota2014,iSpanJi,Chen2018}. 

\subsection{An Application of Graph Trimming} 
Detecting the strongly connected components in directed graphs, the so-called SCC decomposition, is one of the fundamental analysis steps {in many applications} such as social networks~\cite{kumar2010structure}, communication networks~\cite{konect:sun_2016_49561}, knowledge networks~\cite{konect:dbpedia2}, and model checking graphs~\cite{hojati1993bdd}. 
Given a directed graph $G=(V, E)$, a \emph{strongly connected component} of $G$ is a maximal set of vertices $C \subseteq V$ such that every two vertices $u$ and $v$ in $C$  are reachable from each other. 
The early SCC algorithms are based on depth-first search (DFS)~\cite{tarjan1972depth,cormen2009introduction}. 
However, lexicographical-first DFS is \mbox{P-complete} and even the random DFS is hard to parallelize~\cite{Reif1985,aggarwal1988random}. 
The breadth-first search (BFS) based Forward-Backward (FW-BW) algorithm has been proposed.
Unlike DFS, BFS can be parallelized without difficulty. 
Starting from a selected pivot vertex, FW-BW performs a forward BFS to identify the vertex set FW that the pivot can reach, followed by a backward BFS to identify the set BW that can reach the pivot. The intersection between FW and BW is an SCC that contains the pivot~\cite{fleischer2000identifying}. In the worst case, each vertex can be selected as a pivot to travel the whole graph in $\mathcal{O}(m)$, which yields a quadratic time complexity of $\mathcal{O}(mn)$~\cite{fleischer2000identifying}. 
In~\cite{Coppersmith,Fleischer2007}, the worst-case time complexity is improved to $\mathcal{O}(m\log n)$ by using a divide-and-conquer approach.

Interestingly, real-world graphs demonstrate SCC features that follow the \emph{power-law property}~\cite{hong2013fast}, that is, several large SCCs take the majority of vertices and the rest of them are trivial SCCs. More importantly, most of the trivial SCCs are \mbox{size-$1$} SCCs. The key observation is that a \mbox{size-$1$} SCC is easy to identify: it has zero incoming edges or zero outgoing edges. Therefore, graph trimming can be used to remove such \mbox{size-$1$} SCCs in parallel with less computational effort than FW-BW and thus in practice can speed up FW-BW. Analogously to \mbox{size-$1$} SCCs, \mbox{size-$2$}~\cite{hong2013fast} and \mbox{size-$3$}~\cite{iSpanJi} SCCs also can be trimmed but with more computational effort. 

\begin{figure}[htb]
\centering
\includegraphics[scale=0.6]{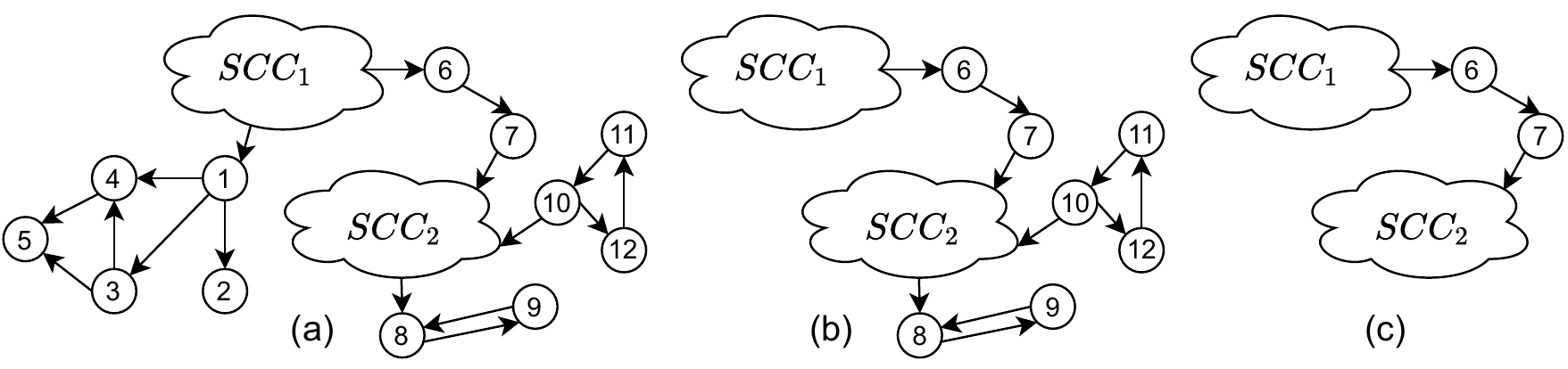}
\caption{A graph that can use graph trimming to remove \mbox{size-$1$}, \mbox{size-$2$} and \mbox{size-$3$} SCCs.}
\label{Figure:example-graph}
\end{figure}

Figure~\ref{Figure:example-graph} illustrates the FW-BW algorithm with graph trimming. Figure~\ref{Figure:example-graph}(a) shows that there are altogether two large SCCs, $SCC_1$ and $SCC_2$ (whose sizes are greatly larger than~$3$) and the other trivial SCCs. It is easy to see that vertices $v_1$ to $v_7$ are \mbox{size-$1$} SCCs, vertices $v_8$ and $v_9$ compose one \mbox{size-$2$} SCC, and vertices $v_{10}$ to $v_{12}$ compose one \mbox{size-$3$} SCC. 
In Figure~\ref{Figure:example-graph}(b), we first try to trim all \mbox{size-$1$} SCCs: (1) in the first repetition, vertices $v_5$ and~$v_2$ are removed since they have no outgoing edge, which causes vertex $v_4$ to have no outgoing edges; (2) in the second repetition, vertex~$v_4$ is removed, which causes vertex $v_3$ to have no outgoing edges; (3) in the third repetition, vertex $v_3$ is removed, which causes vertex $v_1$ to have no outgoing edges;  
(4) in the final repetition, vertex $v_1$ is removed. 
Similarly, in Figure~\ref{Figure:example-graph}(c), \mbox{size-$2$} and \mbox{size-$3$} SCCs can also be removed. Note that vertices $v_6$ and $v_7$, located between two large SCCs, are \mbox{size-$1$} SCCs, but they can not be directly trimmed. 
After the first round of graph trimming, the FW-BW algorithm can identify two large SCCs, $SCC_1$ and $SCC_2$, which also can be deleted from the graph. After removing the two large SCCs, the second round of trimming can remove vertices $v_6$ and $v_7$ with two iterations.

The naive trimming method as used in FW-BW \cite{Fleischer2007} has a quadratic time complexity of $\mathcal{O}(mn)$ in the worst case. The drawback of such trimming is that it sacrifices the better worst-case time complexity of FW-BW, $\mathcal{O}(m \log n)$ \cite{Fleischer2007}.
From our experiments, we noticed that the running time of such trimming will dramatically increase with the number of peeling steps $\alpha$ because of the increasing number of repetitions. 
This is why FW-BW with trimming as in~\cite{hong2013fast} is only efficient for \emph{small-word graphs}. The small-world property states that the \emph{diameters} (greatest shortest-path distance between any pair of vertices) of graphs are very small even for very large graph instances~\cite{watts1998collective}, which always implies a small number of peeling steps. 
The focus of this paper is to improve traditional graph trimming so that algorithms based on FW-BW with trimming~\cite{McLendon2005,hong2013fast,Slota2014,iSpanJi,Chen2018} can be more efficient, especially for non-small-world graphs.

For instance, in~\cite{iSpanJi}, the parallel SCC decomposition algorithm ISPAN is proposed. It combines the power of graph trimming and FW-BW, both of which can be efficiently parallelized. In particular, graph trimming is used at two places; before large SCC detection, trimming is used to remove the \mbox{size-$1$} SCCs; after the large SCC is detected, trimming is again used to remove \mbox{size-$1$}, \mbox{size-$2$}, and \mbox{size-$3$} SCCs. The evaluation uses 56 workers over 16 graphs and shows that ISPAN achieves a significant speedup of 171 - 6591 times  over the sequential DFS-based Tarjan's algorithm~\cite{cormen2009introduction} and of 85 - 1475 times over the parallel DFS-based UFSCC algorithm~\cite{bloemen2016multi}.

\subsection{The New Method} 
Essentially, graph trimming is a kind of \emph{Constrain Satisfaction Problem} (CSP), that is, a set of vertices must satisfy a number of constraints or limitations, e.g.~each vertex needs at least one outgoing edge or it will be removed as a size-1 SCC. 
Many filtering algorithms~\cite{dib2010arc} have been proposed to remove values that obviously do not belong to the solution of a CSP and thus reduce the search space. 
The closest related filtering algorithms to graph trimming are \emph{Arc-Consistency} (AC) algorithms for binary CSPs, in particular 
AC-3~\cite{mackworth1985complexityAC3}, AC-4~\cite{mohr1986arcAC4}, and AC-6~\cite{Bessiitre1999}. 

Table~\ref{Table:ac} summarizes the time and space complexities of these three algorithms. AC-6 has the best worst-case time and space complexities. AC-4 and AC-6 have the same time complexity. However, in reality, AC-3 and AC-6 perform sometimes better than AC-4 due to AC-4 always running close to its worst-case time. The details are explained in the next section.
\begin{table}[htb]
\centering
\begin{tabular}{c | c c }
\hline\noalign{\smallskip}
Algorithm & Time ($\mathcal{O}$) & Space ($\mathcal{O}$) \\ 
\noalign{\smallskip}\hline\noalign{\smallskip}
    AC-3  &    $ed^3$      &     $e + kd$          \\ 
    AC-4     &    $ed^2$      &   $ed^2$        \\ 
    AC-6     &    $ed^2$      &    $ed$        \\ 
\noalign{\smallskip}\hline
\end{tabular}
\caption{The worst-case time and space complexities of three arc-consistency algorithms, where $e$ is the number of arcs, $k$ is the number of variables, $d$ is the size of the largest variable's domain.}
\label{Table:ac}
\end{table}

The key observation is that the graph trimming technique in~\cite{McLendon2005} is like {AC-4} (\emph{AC-4-based}). 
Also, the other widely used graph trimming technique in~\cite{hong2013fast,Slota2014,iSpanJi,Chen2018} is like \mbox{AC-3} (\emph{AC-3-based}). 
Stimulated by AC-6, we design a novel graph trimming algorithm (\emph{AC-6-based}). 
Compared to AC-3-based and AC-4-based trimming, our new AC-6-based trimming is more complicated and not easy to parallelize. 
To the best of our knowledge, there exists little work on parallel AC algorithms~\cite{Cooper1992,Kirousis1993}. In this work, we design efficient sequential and parallel versions of AC-6-based trimming for a multi-threaded shared memory architecture. 

Table~\ref{Table:trim} summarizes the complexities of different parallel graph trimming algorithms in the \emph{work-depth} model, where the \emph{work} is the number of operations used by the algorithm and the \emph{depth} is the length of the longest sequential dependence in the computation. 
We can see that all three trimming algorithms have the different parallel depth, and AC-3-based trimming has a smallest depth. AC-3-based trimming has larger worst-case work and time complexities than the other two algorithms. The AC-6-based and AC-4-based algorithms have the same worst-case work and time complexities. We show that AC-6-based trimming traverses fewer edges and uses less space. For example, over all tested graphs in our first experiments, AC-6-based trimming reduces the number of traversed edges 3.3 - 192.5 times compared with AC-4-based trimming and 1.5 - 44 times compared with AC-3-based trimming.

\begin{table*}[htb]
\centering
\small
 
\begin{tabular}{c|c| c c  c }
 \toprule
            &           & \multicolumn{3}{c}{Worst-Case ($\mathcal{O}$)} \\
 Trimming   & On-The-Fly  &    Work & Depth      & Space \\     
 \midrule
AC-3-based  & \cmark    & $\alpha (n+m)$ & $\alpha \Degout $        &   $n$   \\ 
AC-4-based  & \xmark    & $n+m$         & $|Q_p| \Degin \Degout$      &  $n+m$  \\  
AC-6-based  & \cmark    & $n+m$         & $|Q_p| \Degin^2$      &   $n$   \\ 
 \bottomrule
 
\end{tabular}
\caption{The worst-case work, depth, and space complexities of parallel graph trimming algorithms, where $n$ is the number of vertices, $m$ is the number of edges, $\mathcal{P}$ is the number of total workers, $\alpha$ is the number of peeling steps, $\Degout$ is the maximal out-degree for all vertices, $ \Degin$ is the maximal in-degree for all vertices, $|Q_p|$ is the upper-bound size of waiting sets among $\mathcal P$ workers such that sometimes $|Q_p| \geq \alpha$.}
\label{Table:trim}
\end{table*}

To parallelize the AC-4-based and AC-6-based algorithms, 
the conventional way is with mutual exclusion by using \texttt{Lock} and \texttt{Unlock} operations that guarantee exclusive access to data structures shared by multiple workers. 
In this work, however, we use atomic primitives to minimize the synchronization overhead. 

\subsection{On-the-fly Property}
The \emph{on-the-fly} property~\cite{bloemen2016multi} means an algorithm can run on an implicit graph defined as $G=(v_0, \texttt{POST})$, where $v_0$ is the \emph{initial vertex} and $\texttt{POST}(v)$ is a function that returns all of the \emph{successors} of vertex $v$.
One drawback of the FW-BW method is that the backward search requires reverse edges, which means all edges have to be loaded into memory; storing the graph as an adjacent list with only outgoing edges is not sufficient. 
The on-the-fly property is necessary when handling large graphs that occur in e.g.~verification~\cite{merz2000model}, as it may allow the algorithm to terminate early after processing only a fraction of the graph without needing memory space for loading the whole graph. It also benefits algorithms that rely on implicit graphs~\cite{pelanek2007beem}, in which the edges are calculated online by function $\texttt{POST}(v)$. 

The on-the-fly properties of three graph trimming algorithms are summarized in Table \ref{Table:trim}. It is easy to see that the AC-4-based trimming cannot run on-the-fly as it requires reverse edges and thus the whole graph must be loaded into the memory. AC-3-based and AC-6-based trimming can run on-the-fly as they only rely on the post of vertex $v$ when traversing each vertex $v\in V$ and their space usage is bounded by $\mathcal O (n)$. However, compared with AC-3-based trimming, AC-6-based trimming needs much less work. Note that on implicit graphs, all the edges are computed online by the function $\texttt{POST}(v)$, which typically costs more running time than directly loading edges from memory like with explicit graphs. The proposed AC-6-based trimming traverses fewer edges than AC-3-based trimming and thus performs better on implicit graphs. 

\subsection{Contribution}
The contributions of this work are summarized below:
\begin{itemize}
    \item We provide a formal definition of graph trimming based on the Constraint Satisfaction Problem (CSP) and Arc-Consistency (AC). Following three well-known arc-consistency algorithms, that is, AC-3, AC-4, and AC-6, we categorize the existing graph trimming algorithms as AC-3-based~\cite{hong2013fast,Slota2014,iSpanJi,Chen2018} and AC-4-based algorithms~\cite{McLendon2005}.
    
    \item We revisit the existing parallel AC-3-based algorithm. We give the detailed steps of the AC-4-based algorithm and parallelize it using atomic primitives.
    
    \item We propose a novel AC-6-based algorithm that has optimized time and space complexities. We further parallelize the AC-6-based algorithm using atomic primitives. 
    These are the main contributions of this work.  
    
    \item For all three graph trimming algorithms, we formally discuss their correctness, time complexity, and space complexity. The time complexities for parallel algorithms are analyzed in the work-depth model.
    
    \item Finally, for all three parallel trimming algorithms, our experiments compare the number of traversed edges and practical running time with 1 to 16 workers over a variety of real and synthetic graphs.
\end{itemize}

This paper is organized as follows. Section~2 provides the background. Section~3 discusses the relation between graph trimming and Arc Consistency. 
Section~4 revisits the AC-3-based graph trimming algorithm. 
Section~5 provides the AC-4-based graph trimming algorithm. 
Section~6 proposes the AC-6-based graph trimming algorithm. 
The related work is discussed in Section~7. 
In Section~8, we discuss the implementations. 
In Section~9, we provide the experimental evaluation over a variety of data graphs.
Section~10 concludes three trimming algorithms and discusses the future work.

\section{Preliminaries}
Given a directed graph $G=(V,E)$, let $n = |V|$ and $m = |E|$ be the numbers of vertices and edges, respectively. A vertex $v$ in graph $G$ is also denoted as $v(G)$. As opposed to an undirected graph, $(v,w)\in E$ does not imply that $(w,v)\in E$. 
The $\post$ of vertex $v$ in $G$ is the set of all the successors (outgoing edges) of $v$, defined by $v.\post = \{w \mid (v,w)\in E\}$; 
when the context is clear, we use $v.\post$ instead of $v(G).\post$. 
The $\pre$ of vertex $v$ is the set of all the predecessors (ingoing edges) of~$v$, defined by $v.\pre=\{w \mid (w, v)\in E\}$. 
For each vertex $v\in V$, its \emph{out-degree} is the number of  successors $|v.\post|$ and its \emph{in-degree} is the number of predecessors $|v.\pre|$.
To analyze the time complexity of trimming algorithms, we use $ \Degout$ and $\Degin$ to denote the maximum out-degree and in-degree among all vertices in a graph $G$, respectively.

A transposed graph $G^T = (V, E^T)$ is equivalent to the graph $G = (V, E)$ with all its edges reversed, $E^T = \{(w, v) \mid (v, w) \in E\}$. It is easy to see that $v(G).\post = v(G^T).\pre$ and $v(G).\pre=v(G^T).\post$ for each $v\in V$. A transposed graph $G^T$ can be generated in order to efficiently obtain $v(G).\pre$ without traversing the whole original graph $G$.

\begin{definition}[Trimmed Graph]
\label{def:TrimmedGraph}
Given a directed graph $G=(V,E)$, the trimmed graph $G' = (V', E')$ with $V'\subseteq  V$ and $E'\subseteq E$ is a maximal subgraph of $G$, where each vertex has at least one outgoing edge, formally $\forall v \in V': v.\post \neq \emptyset$.
\end{definition}

This work focuses on the graph trimming algorithms that can obtain trimmed graphs according to~Definition \ref{def:TrimmedGraph}.
Without changing the original graph $G=(V,E)$, each vertex $v\in V$ is assigned a status, denoted as $\it v.\status$, with values {\LIVE} and {\DEAD}, which indicates if vertex $v$ is located in the graph (live) or removed (dead), respectively.

\subsection{Graph Storage}
In this work, explicit graphs and implicit graphs are discussed. Explicit graphs are typically stored in the \emph{compressed sparse row} (CSR) format~\cite{Hong2012,hong2013fast}. 
This format uses two arrays to represent the graph: an $\mathcal{O}(n)$-sized array stores an index to the beginning of each vertex's adjacency list and an $\mathcal{O}(m)$-sized array stores each vertex's adjacency list. The CSR representation is compact, memory bandwidth-friendly, and thus suitable for efficient graph traversals. It is easy to see that successors of each vertex $v\in V$ are ordered and thus can be traversed one by one in order.

On modern computers, registers can move data around in single clock cycles. However, registers are very expensive. The dynamic random access memory is very cheap but takes hundreds of cycles after a request to receive the data. To bridge this gap between them are the cache memories, named L1, L2, L3 in decreasing speed and cost. If the data is stored in memory sequentially, the CPU can prefetch the data into the cache for fast accessing, which is cache-friendly. For a graph stored in CSR format, we can see that sequentially traversing all edges is cache-friendly as the cache hit rate is high, but randomly traversing all edges is not cache-friendly as the cache hit rate is low.

Implicit graphs are defined as $G=(v_0, \texttt{POST})$ assuming that all the vertices in $G$ are reachable from vertex $v_0$, where $v_0$ is the \emph{initial vertex} and $\texttt{POST}(v)$ is a function that returns all of the \emph{successors} of vertex $v$, that is, $\texttt{POST}(v)=v.\post$. One kind of implicit graphs are model checking graphs \cite{pelanek2007beem} such that for each vertex $v$ in a graph $G$, all the edges are calculated online by $\texttt{POST}(v)$.
Another kind of implicit graphs are external graphs such that all the edges are stored on disks sequentially; once a vertex $v$ is traversed, the edges of $v$ are loaded into memory. The advantage of implicit graphs is that they allow handling large graphs with limited memory usage. However, much running time is spent on generating the edges via $\texttt{POST}(v)$. If an algorithm can run on implicit graphs without loading the whole graphs into memory, we say this algorithm has the \emph{on-the-fly property}


\subsection{Constraint Satisfaction Problem} 
A \emph{constraint satisfaction problem} (CSP)~\cite{Russell2010AI,dib2010arc} can be defined as a triple $P=(X, D, C)$, where $X=\{X_1,...,X_n\}$ is a set of $n$ variables, $D = \{D(X_1), ... ,D(X_n)\}$ is the set of $n$ domains such that $D(X_i)$ is a set of possible values of variable $X_i$, and $C$ is a set of constraints that specify allowable combinations of values. A \emph{solution} of a constraint set $C$ is an instantiation of the variables such that all constraints are satisfied.
Here, we restrict to \emph{binary constraints} $C_{ij}$ between pairs $(X_i, X_j)$ of variables, i.e.~$C=\{C_{ij} \mid i,j\in 1...n\}$.

\subsection{Arc-Consistency} 
 A value $v_i\in D_i$ is \emph{binary consistent} with a constraint $C_{ij}$ if there exists $v_j \in D_j$ such that $(v_i, v_j)$ satisfies $C_{ij}$. Then $v_j \in D_j$ is called a \emph{support} of $v_i \in D_j$ over $C_{ij}$. A value $v_i \in D_i$ is \emph{viable} if it has supports for every $D_j$ such that each $C_{ij}\in C$ is satisfied.  
A variable in a CSP is \emph{arc-consistent} (AC) if every value in its domain satisfies each binary constraint $C_{ij}\in C$.

Several AC algorithms have been proposed for removing values that are not viable.
AC-1~\cite{mackworth1977consistencyAC1-2} revisits all the binary arcs that have to be revisited once some domains are reduced. Improving on AC-1, algorithm AC-2~\cite{mackworth1977consistencyAC1-2} only revisits the arcs that are affected by reducing some domains. Algorithm \mbox{AC-3}~\cite{mackworth1977consistencyAC1-2,mackworth1985complexityAC3} generalizes and simplifies AC-2.

Algorithm~\ref{Alg:AC-3} shows the detailed steps of AC-3. 
Initially, the global set $Q$ includes all constraint arcs $C_{ij}\in C$ (line 1). Each constraint arc $C_{ij}$ is picked and then removed from the set $Q$ (line 3), and each pair of values in the domains $D(X_i)$ and $D(X_j)$ are checked by the procedure \texttt{Revise} (line 4), that is,
for each value $v_i\in D(X_i)$, if $D(X_j)$ does not contain a value $v_j$ such that $(v_i, v_j)$ satisfies the constraint $C_{ij}$, the value $v_i$ is repeatedly removed from $D(X_i)$ (lines 7 - 13). 
If $D(X_i)$ is changed, the associated constraints $C_{ij}$ are placed into $Q$ again (lines~5 and~6). 
This process is repeated until the set $Q$ becomes empty (line 2).
It is easy to see that AC-3 is not efficient since the revision of any domain will force neighbor constraints to be revisited again.

\begin{algorithm}[hbt]
\DontPrintSemicolon
\SetAlgoLined
\SetAlgoNoEnd
\SetKwData{True}{TRUE}
\SetKwData{False}{FALSE}
\SetKwFunction{Revise}{Revise}
\SetKwInOut{Input}{input}\SetKwInOut{Output}{output}
    \Input{An arc-consistency problem $P=(X,D,C)$}
    \Output{A filtered domain set $D$}
$Q := C$\;
\While{$Q \neq \emptyset$}{
    remove a constraint $C_{ij}$ from $Q$\;
    \If{\Revise{$X_i, X_j, D$}}{
        \For{$X_k \in \{X_{k'}: C_{k'i} \in C\}\setminus \{X_j\}$ }{
            $Q := Q \cup \{C_{ki}\}$
        }
    }
}
\medskip
\SetKwProg{myproc}{procedure}{ \Revise{$X_i, X_j, D$}}{}
\myproc{}{
            $\revised := \False$\;
    \For{$v_i\in D(X_i)$}{
        \If{\rm{no value $v_j$ in $D(X_j)$ satisfies $C_{ij}$}}{
            delete $v_i$ from $D(X_i)$\;
            $\revised := \True$
        }
    }
    \Return $\revised$\;
}

\caption{AC-3}
\label{Alg:AC-3}
\end{algorithm}

AC-4~\cite{mohr1986arcAC4} improves the worst-case time complexity of AC-3 by using auxiliary data structures, \emph{supports} and \emph{counters}, but its average running time is close to the worst-case time complexity. 
However, AC-3 has better average running time and space usage than AC-4 and thus AC-3 is always preferred to \mbox{AC-4}~\cite{Russell2010AI} in practice.
Algorithm AC-6~\cite{Bessiere1994AC6} combines AC-3 and AC-4. It only records one support for each value, unlike AC-4 which records all supports, since a single  support is enough to prove that a value is viable. Because of this, AC-6 has the same worst-case time complexity as AC-4 but averagely performs much better than AC-4 in many applications. Further, AC-6 needs less space than AC-4 since for each value only a single support is recorded. 
The corresponding time, work, and space complexities in the worst case are summarized in Table~\ref{Table:ac}.

\subsection{Parallel Complexity Analysis} 
\label{sec:complexity}
We assume parallel programs run on shared-memory
multi-core machines, where different cores access a shared global memory simultaneously. Shared-memory parallelism has many advantages, but writing correct, efficient, and scalable shared-memory multi-core programs is difficult.
In this paper, the parallel graph trimming algorithms are designed for nested \emph{fork-join} parallelism, in which a \emph{fork} specifies workers that can execute in parallel, and a \emph{join} specifies a synchronization point among multiple workers. In practice, our parallel algorithms can be implemented by OpenMP \cite{dagum1998openmp}.

We analyze our parallel algorithms in the \emph{work-depth} model~\cite{cormen2009introduction,shun2017shared}, where the \emph{work}, denoted as $\mathcal W$, is the total number of operations that are used by the algorithm and the \emph{depth}, denoted as $\mathcal D$, is the longest length of sequential operations \cite{jeje1992introduction}. 
This model is particularly convenient for analyzing nested parallel algorithms. 
Assuming that a scheduler dynamically load-balances a parallel computation across all available workers, the expected running time is $\mathcal{O}(\mathcal W/\mathcal{P} +\mathcal{D})$ when using $\mathcal{P}$ workers. For the multi-core architecture, a worker is a working process corresponding to a physical core.
In particular, for sequential algorithms, the work and the depth terms are equivalent. 
A parallel algorithm is \emph{work-efficient}
if its work is asymptotically equal to the work of the fastest sequential algorithm for the same problem \cite{blelloch2010parallel}. 

\begin{definition}[Number of Peeling Steps $\alpha$]
\label{def:d}
Given a directed graph $G=(V, E)$, integer $\alpha$ is defined as the number of peeling steps: in the peeling process, a step removes all vertices with zero out-degrees and thus decrements the out-degrees of adjacent neighbors. Neighbors whose out-degrees becomes zero must be removed in the next step. This is repeated until no vertices have an out-degree of zero.
\end{definition}

To analyze the depth of our trimming algorithms, the number of peeling step $\alpha$ is introduced in Definition \ref{def:d}, which is analogous to the \emph{peeling-complexity} proposed to analyze the depth of parallel $k$-core decomposition algorithms \cite{dhulipala2017julienne}. 
Essentially, $\alpha$ is a property of graphs, which indicates the longest chain size of trimmable vertices. It is easy to see that $\alpha$ can be as large as $n$ in the worst case, e.g.~in a chain graph. However, $\alpha$ is significantly smaller than $n$ in practice. For example, for the graphs in our experiments, $\alpha$ ranges from $3$ to $11,686$, which is small compared to their millions of vertices.

\subsection{Atomic Primitives}
\label{sec:atomic}
All algorithms are implemented for shared-memory parallel machines; that is, multiple workers access the same memory \cite{shun2017shared}.
The conventional way is with mutual exclusion by using \texttt{Lock} and \texttt{Unlock} operations that guarantee exclusive access to data structures shared by multiple workers. 
Compared with using locks, a implementation by using atomic primitives has much less synchronization overhead and the unexpected delay while workers are within critical sections can be highly reduced. The compare\&swap (\texttt{CAS}) and fetch\&add (\texttt{FAA}) operations are universal atomic primitives that are supported on the majority of current parallel architectures~\cite{valois1995lock,michael2002high,milman2018bq}.

\iffalse
The \texttt{CAS} operation is a universal atomic primitive that is supported on the majority of current worker architectures~\cite{valois1995lock,michael2002high,milman2018bq}.
\iffalseAs shown in Algorithm~\ref{Alg:cas}, \texttt{CAS} takes three arguments, a variable (location) $x$, an old value $a$ and a new value $b$. It atomically checks the value of the variable, and if it equals the old value, it updates the pointer to the new value and then returns \textit{true}, otherwise, it returns \textit{false} to indicate it did not succeed. Angular brackets, $\langle ... \rangle$, are used to indicate atomicity.
The support lists are shared by multiple workers but can be implemented lock-free. We can choose concurrent single-linked lists proposed in~\cite{valois1995lock}, as only the \texttt{CAS} primitive is needed.
\fi 

 As shown in Algorithm~\ref{Alg:cas}, the \texttt{CAS} atomic primitive takes three arguments, a variable (location) $x$, an old value $a$ and a new value $b$. It checks the value of the variable $x$, and if it equals to the old value $a$, it updates the pointer to the new value $b$ and then returns \textit{true}; otherwise, it returns \textit{false} to indicate that the updating fails. Here, we use a pair of \emph{angular brackets}, $\langle ... \rangle$, to indicate that the operations in between are executed atomically.
 
\begin{algorithm}[hbt]
\SetKwData{True}{\small TRUE}
\SetKwData{False}{\small FALSE}
\caption{CAS($x, a, b$)}
\label{Alg:cas}
\DontPrintSemicolon
\SetAlgoLined
\SetAlgoNoEnd
    $\langle\,$\If{$x = a$}{
        $x := b$; \Return \True
    }
    \lElse{
        \Return \False $\rangle$ 
        \textcolor{darkgreen}{\tcc*[f]{$\langle ... \rangle$ atomic}} 
    }
\end{algorithm}

The \texttt{FAA} atomic primitive is shown in Algorithm \ref{Alg:faa}. The old value of $x$ is fetched and added by $a$. The new value of $x$ is returned. For instance, there is a race condition when one worker is executing $``x := x + a"$ and the other worker is executing $``x := x + b"$ concurrently.  Using \texttt{FAA} can efficiently get the correct result without workers affecting each other.

\begin{algorithm}[hbt]
\SetKwData{True}{\small TRUE}
\SetKwData{False}{\samll FALSE}
\caption{FAA($x, a$)}
\label{Alg:faa}
\DontPrintSemicolon
\SetAlgoLined
\SetAlgoNoEnd
    $\langle x:=x+a\rangle$  \textcolor{darkgreen}{\tcc*[f]{$\langle ... \rangle$ atomic}} 
\end{algorithm}



\section{Graph Trimming as Arc Consistency}
Intuitively, we can regard graph trimming as a graph with a constraint that each vertex has at least one outgoing edge. Based on this observation, we define graph trimming as an arc-consistency problem with one single variable, viz.~the set of all vertices, and a single binary constraint, viz.~each vertex must have at least one outgoing edge as one support. Then, trimming a graph means determining the domain of available vertices. 

More formally, given a directed graph $G=(V, E)$, graph trimming can be defined as an arc-consistency problem $(X,D,C)$ with variables $X=\{X_1\}$, domains $D=\{D(X_1)\}$ and constraint $C=\{C_{11}\}$.
Here, we assume that $X_1=V$ and $C_{11} = E$, that is, each vertex $v_1 \in D(V)$ must has at least one support vertex $v_1' \in D(V)$ in the same domain, where $(v_1, v_1') \in E$. 

Consequently, three important AC algorithms, AC-3, AC-4, and AC-6, can be applied to graph trimming. Interestingly, we find that one widely used graph trimming method~\cite{hong2013fast,Slota2014,iSpanJi,Chen2018} is analogous to AC-3 (we call it AC-3-based). The other~\cite{McLendon2005} is analogous to AC-4 (we call it AC-4 based); however, the detailed steps are not discussed, and a parallel version is not provided. As a contribution, we design a novel graph trimming algorithm based on AC-6 (we call it AC-6-based). 

In Table \ref{tb:notations}, we summarize the notations that will be frequently used when discussing the graph trimming algorithms. 
\begin{table}[htb]
\centering
\begin{tabular}{l|l}
\toprule
Notation &  Description\\
\midrule
$G=(V,E)$              &  a directed graph with $n$ vertices and $m$ edges\\
$G^T=(V, E^T)$           & a transposed graph of $G=(V,E)$\\
$u(G).\degin$             &  the in-degree of $u$ in $G$\\ 
$u(G).\degout$            &  the out-degree of $u$ in $G$\\ 
$u(G).\post$        & the successors of $u$ in $G$\\
$u(G).\pre$         & the predecessor of $u$ in $G$\\
$u(G).S$ & the supporting set used by AC-6-Trimming\\ 
$u(G).\status$ & the status ({\LIVE} or {\DEAD}) of $u$ in $G$\\
$\Degin$             &  the maximal in-degree among all vertices in $G$\\ 
$\Degout$            &  the maximal out-degree among all vertices in $G$\\ 
$\alpha$ &  the number of peeling steps in $G$\\ 
$Q$ & the waiting set used by AC-4-Trimming and AC-6-Trimming\\
$\mathcal P$ & the total number of workers \\ 
$Q_p$ & a private waiting set used by workers $p$\\
$|Q_p|$ & the upper-bound size of waiting sets among $\mathcal P$ workers\\
\bottomrule
\end{tabular}
\caption{The notations that frequent used when discussing the graph trimming algorithms.}
\label{tb:notations}
\end{table}

\section{AC-3-Based Graph Trimming}
In the graph trimming problem, there exists only a single variable and a single constraint. Therefore, AC-3, as shown in Algorithm \ref{Alg:AC-3}, can be simplified when applied to graph trimming. The idea is straightforward: (1) for all vertices in a graph, the vertices with zero out-degrees are removed; (2) this process is repeated until the graph does not change. This naive trimming method is widely used~\cite{hong2013fast,Slota2014,iSpanJi,Chen2018} for quickly removing the \mbox{size-$1$} SCCs, but the correctness and complexities are not formally discussed.
In this section, we revisit the existing parallel AC-3-based algorithm for graph trimming and formally discuss the correctness and complexities. The sequential AC-3-based algorithm is immediate and not discussed further.

\subsection{The Parallel AC-3-Based Algorithm }
Algorithm~\ref{Alg:TraditionalTrim} shows the detailed steps of the parallel AC-3-based algorithm for graph trimming. 
The procedure $\texttt{ZeroOutDegree}(v)$ (lines 11 - 14) returns \TRUE~if vertex $v$ has at least one available outgoing edge and \FALSE~otherwise. 
All vertices in $V$ are initialized as $\texttt{LIVE}$.
After partitioning $V$ into $V_1\dots V_{\mathcal{P}}$, we have $\mathcal{P}$ workers execute the procedure $\texttt{Trim}_p(V_p)$ in parallel (lines 4 and 5).
The main procedure $\texttt{Trim}_p(V_p)$ (lines 7 - 10) removes the vertices in $V_p$ that have an out-degree of zero. 
The removing process repeats until the graph does not change (lines 2 - 6). One advantage of this algorithm is that it is easy to parallelized: each copy of procedure $\texttt{Trim}_p(V_p)$ for a worker $p$ (line 5) can run in parallel with only \emph{change} as the sole shared variable.


\begin{algorithm}[hbt]
\DontPrintSemicolon
\SetKwData{Live}{\small LIVE}
\SetKwData{Dead}{\small DEAD}
\SetKwData{False}{\small FALSE}
\SetKwData{True}{\small TRUE}
\SetKw{With}{with}
\SetKwFunction{Indegree}{InDegree}
\SetKwFunction{Outdegree}{ZeroOutDegree}
\SetKwFunction{Trim}{Trim$_p$}
\SetKwFunction{TrimOne}{Trim$_1$}
\SetKwFunction{TrimP}{Trim$_{\mathcal{P}}$}
\SetAlgoLined
\SetAlgoNoEnd
\SetKwInOut{Input}{input}\SetKwInOut{Output}{output}
\Input{Graph $G=(V, E)$}
\Output{Trimmed Graph $G$}
\lFor{$v\in V$}{$v.\status := {\Live}$}
\Repeat{$\neg~\change$}{
    $\change := \False$\;
    partition $V$ into $V_1 \dots V_{\mathcal{P}}$\;
    \TrimOne{$V_1$}$\parallel \dots \parallel$ \TrimP{$V_{\mathcal{P}}$}\;
}
\medskip
\SetKwProg{myproc}{procedure}{ \Trim{$V_p$}}{}
\myproc{}{
        \For{$v \in V_p: v.\status = {\Live}$}{
            \If{$\Outdegree{v}$}{
                $v.\status :={\Dead}$; $\change := \True$}
        }
}
\medskip

\SetKwProg{myproc}{procedure}{ \Outdegree{$v$}}{}
\myproc{}{
    \For(){$w\in v.\post$ \With $w.\status = \Live$}{
        \Return \False
    }
    \Return \True
}
 

\caption{Parallel AC-3-based Graph Trimming}
\label{Alg:TraditionalTrim}
\end{algorithm}


For the specific implementations in~\cite{hong2013fast,iSpanJi,Chen2018}, there are two strategies to improve the AC-3-based algorithm for graph trimming.
\begin{itemize}
\item If the transposed graph $G^T$ is loaded in memory, another constraint can be considered, that each vertex $v \in V$ must have at least one available incoming edge. That means the in-degree also be checked (line 9). In this case, more \mbox{size-1} SCCs can be quickly trimmed. The problem is that the transposed graph is required, which costs $\mathcal O(n+m)$ memory space.

\item The number of repetitions can be limited to a constant number like 3 or the repetitions stop when the number of removed vertices is less than a threshold like 100 (line~6). The problem is that some of the trimable vertices may not be removed. This strategy is sometimes effective at reducing the computational time but sometimes not, since the worst-case time complexity is not improved. 
\end{itemize}

\paragraph{Correctness.}
For the correctness, trimming has to be sound and complete. 
Soundness means that all removed vertices, which are assigned a status of \texttt{DEAD}, must have no outgoing edges or only edges to removed vertices:
\begin{equation} \begin{split} \label{eq:sound}
sound(V)~\equiv~\forall v \in V:~& v.\status = {{\DEAD}} \implies \\
&(\forall w \in v.\post: w.\status = \texttt{DEAD})
\end{split} \end{equation}
Completeness means that all vertices that have no outgoing edges or have only outgoing edges to removed vertices are removed:
\begin{equation} \begin{split} \label{eq:complete}
complete(V)~\equiv~\forall v \in V:~& (\forall w \in v.\post: w.\status = \texttt{DEAD}) \implies \\
&v.\status =\texttt{DEAD}
\end{split} \end{equation}
The algorithm has to ensure both soundness and completeness for all vertices in the graph:
\begin{equation} \label{eq:soundcomplete}
    sound(V)~\land~complete(V)
\end{equation}
which is equivalent to:
\begin{equation} \begin{split}
\forall v \in V: v.\status = \texttt{DEAD}~\equiv~(\forall w \in v.\post: w.\status = \texttt{DEAD})
\end{split} \end{equation}



For arguing about the correctness of for-loops, we use following rule: consider the loop $\bf for~\it x \in X~\bf do~\it S$ and let $P(X')$ be a predicate. If (1) initially $P(\emptyset)$ holds and (2) under precondition $P(X')$ the body $S$ establishes postcondition $P(X' \cup \{x\})$ for any $X' \subset X$ and $x \in X \setminus X'$, then finally $P(X)$ holds; $X'$ is the set of visited elements and $P(X')$ is the loop invariant.

\begin{theorem}[Soundness]
For any $G=(V,E)$ Algorithm~\ref{Alg:TraditionalTrim} terminates with $sound(V)$.
\end{theorem}
\begin{proof}
The invariant of the for-loop of $\texttt{Trim}_p$ (lines 8-10) is $\it sound(V')$: initially that holds as the universal quantification in $\it sound(V')$ is empty. The invariant is preserved as $v.\status$ is only set to \texttt{DEAD} if the status of all $w \in v.\post$ is \texttt{DEAD} (lines 9 and 10). We use the fact that $\texttt{ZeroOutDegree($v$)}$ returns $(\forall w \in v.\post : w.\status = \texttt{DEAD})$. The postcondition of $\texttt{Trim}_p\texttt{(}V_p\texttt{)}$ is therefore $\it sound(V_{\it p})$. The postcondition (line 5) is then $sound(V_1)\land \dots \land sound(V_{\mathcal{P}})$, which is equivalent to $sound(V)$. Thus $sound(V)$ is the invariant of the repeat-until loop (lines 2 - 6) and therefore holds on termination.
\qed \end{proof}

\begin{theorem}[Completeness]
For any $G=(V,E)$ Algorithm~\ref{Alg:TraditionalTrim} terminates with $complete(V)$.
\end{theorem}
\begin{proof}
The invariant of the for-loop of $\texttt{Trim}_p$ (lines 8-10) is $\it \neg change \implies complete(V')$, where $V'$ is the set of visited vertices.
If $\it change$ is $\texttt{TRUE}$, the invariant is obviously preserved as $\it change$ is not set to $\texttt{FALSE}$ in this or any other parallel copy of $\texttt{Trim}_p$. 
Suppose $\it change$ is $\texttt{FALSE}$ and $complete(V')$ holds.
For $v\in V\setminus V'$ that remains $\texttt{LIVE}$, the procedure \texttt{ZeroOutDegree}$(v)$ (line 9), which computes $(\forall w \in v.\post : w.\status = \texttt{DEAD})$, must return false. 
Since setting $\it v.\status$ to $\texttt{DEAD}$ may invalidate $complete(V' \cup \{v\})$ for this or some other parallel copy of $\texttt{Trim}_p$, variable $\it change$ is set to \texttt{TRUE}, which re-establishes the invariant for this and all other parallel copies of $\texttt{Trim}_p$.
\qed \end{proof}

\paragraph{Complexities.}
The complexity of parallel AC-3-based graph trimming has been discussed in existing work ~\cite{hong2013fast,Slota2014,iSpanJi,Chen2018}, but not with the work-depth model. We adopt the work-depth model to analyze the time complexity.

\begin{theorem}
Algorithm \ref{Alg:TraditionalTrim} requires $\mathcal O(\alpha(n+m))$ expected work, $\mathcal{O}(\alpha\Degout)$ depth, and thus $\mathcal{O}(\alpha(n+m)/\mathcal{P}+\alpha\Degout)$ time complexity. 
\end{theorem}
\begin{proof} \rm
For the inner for-loop (lines 8 - 10), checking the out-degree of each vertex $v\in V$ requires $\mathcal{O}(n+m)$ work in the worst case since all edges may need to be traversed in case of some vertices are removed. For the outer repeat-loop (lines 2 - 6), all vertices must be checked again once at least one vertex is removed. The repetition is carried out $\alpha$ times. Therefore, the expected work is $\mathcal O(\alpha(n+m))$. 

We analyze the working depth. For the procedure \texttt{Trim$_p$}, the inner for-loop (lines 12 and 13) in procedure \texttt{ZeroOutDegree} run in sequential with depth $\mathcal O(\Degout)$. The repetition (lines 2 - 6) is carried out $\alpha$ times. Therefore, the theoretical working depth is $\mathcal O(\alpha\Degout)$, and thus the theoretical time complexity is $\mathcal O(\alpha(n+m)/\mathcal P + \alpha\Degout)$.
\qed \end{proof}


\begin{theorem}
The space complexity of Algorithm \ref{Alg:TraditionalTrim} is $\mathcal O(n)$.
\begin{proof} \rm
Each vertex $v\in V$ requires $\status$ in memory to record if vertex $v$ is \LIVE~or \DEAD. Besides $\status$, no other auxiliary data structures are utilized. Therefore, the space complexity is $\mathcal O(n)$.
\qed \end{proof}
\end{theorem}

\section{AC-4-Based Graph Trimming}
AC-4 improves the worst-case time complexity of AC-3 by using auxiliary data structures, \emph{supports} and \emph{counters}. Specifically, for each value in its domain, its supports are recorded, and its total number of supports is recorded with a counter. When removing one value, the corresponding counters located by supports are decreased by one. The values whose counters are reduced to zero must be removed, which may cause other values to be removed.
In a word, the supports and counters are used for efficient propagation after unqualified values are removed.

The AC-4 algorithm can be applied to graph trimming, which we call AC-4-based trimming. For a directed graph $G=(V,E)$, the supports can be simplified as the transposed graph $G^T=(V,E^T)$, and the counters can be implemented by out-degree counters for all vertices $v \in V$, denoted as $v.\degout$. 
AC-4-based graph trimming is used in~\cite{McLendon2005} for quickly removing \mbox{size-$1$} SCCs to speed up SCC decomposition; nevertheless, the details of this algorithm are not discussed, and its parallel version is not given. In this section, we provide both sequential and parallel AC-4-based algorithms.

\subsection{The Sequential AC-4-Based Algorithm}
Algorithm \ref{Alg:seq-ac4trim} shows the detailed steps of the sequential AC-4-based algorithm. Compared to the AC-3-based algorithm, there are two new data structures: 1)~the transposed graph $G^T=(V,E^T)$ is required for accessing the predecessors of a vertex $v\in V$ (line 6); 2)~a waiting set $Q$ is required for propagation when processing removed vertices (line 3).
The procedure \texttt{DoDegree($v, Q$)} (lines 9 - 11) removes the vertex $v$ and puts it into $Q$ for propagation if $v$ is \LIVE~and its out-degree counter $v.\degout$ is zero. That means all vertices in the waiting set $Q$ are dead. 

Now we explain Algorithm~\ref{Alg:seq-ac4trim}. 
Initially, for all vertices, their status and out-degree counters are correctly initialized (line 1). 
For each vertex $v \in V$, the out-degree counter is checked by calling procedure \texttt{DoDegree($v, Q$)} (line 3). The removed vertices are added into the wait set $Q$ and then propagated to update the out-degree counters of other vertices (lines 4 - 8). That is, for each vertex $w\in Q$, all its predecessors' out-degree counters are off by 1 and then checked by the procedure \texttt{DoDegree($v, Q$)} (lines 6 - 8). During this process, new vertices may be removed and added into the waiting set $Q$ so that the algorithm does not terminate until $Q$ becomes empty (line 4).

\begin{algorithm}[!htb]
\caption{Sequential AC-4-based Graph Trimming}
\label{Alg:seq-ac4trim}
\DontPrintSemicolon
\SetAlgoLined
\SetAlgoNoEnd
\SetKwFunction{ContractSCC}{ContractSCC}%
\SetKwFunction{Contract}{Contract}
\SetKwFunction{Next}{DoPost}
\SetKwFunction{Core}{Core}
\SetKwFunction{First}{FirstPost}
\SetKwFunction{Next}{DoDegree}
\SetKwFunction{Indegree}{InDegree}
\SetKwFunction{Outdegree}{OutDegree}
\SetKw{Continue}{continue}
\SetKwData{Live}{\small LIVE}
\SetKwData{Dead}{\small DEAD}
\SetKwData{Gray}{gray}
\SetKwData{Unseen}{unseen}
\SetKwData{Instack}{instack}
\SetKwData{Complete}{complete}
\SetKw{With}{with}
\SetKwInOut{Input}{input}\SetKwInOut{Output}{output}
\Input{Graph $G=(V, E)$ and its transposed graph $G^T=(V,E^T)$}
\Output{Trimmed graph $G$}

\lFor{$v\in V$}{$\it v.\status, v.\degout := {\Live}, |v(G).\post|$}
\For{$v \in V$ \With $v.status = \Live$}{
    $Q := \emptyset;$ \Next{$v, Q$}\;
    \While{$Q \neq \emptyset$}{
        remove a vertex $w$ from $Q$\;
        \For{$v'\in w(G^T).\post$}{
            $v'.\degout := v'.\degout - 1$\;
            \Next{$v', Q$}
        }
    }
}
\medskip
\SetKwProg{myproc}{procedure}{ \Next{$v, Q$}}{}
\myproc{}{
  \If{${v.\degout} = 0 \land v.\status = {\Live}$}{
    $v.\status := {\Dead}; Q:=Q\cup \{v\}$}
}
\end{algorithm}

\iftrue
\paragraph{Correctness.}
We show soundness and completeness together.

\begin{theorem}[Soundness and Completeness]
For any $G=(V,E)$ Algorithm~\ref{Alg:seq-ac4trim} terminates with $sound(V)$ and $complete(V)$.
\end{theorem}
\begin{proof}
Let $V'$ be the set of vertices visited by the outer for-loop (lines 2 - 8).
The invariant of the outer for-loop (lines 2 - 8) is that all vertices are sound, all visited vertices are complete, and that for each vertex $v\in V'$ the counter $v.\degout$ is the number of live vertices of outgoing edges:
\begin{dmath*} 
sound(V)~\land~complete(V')~\land~(\forall v \in V : v.\degout = |\{w \in v.\post \mid w.\status = {\LIVE}\}|)
\end{dmath*}
The invariant holds initially as setting all vertices to {\LIVE} makes them sound and~$V'$ is initially empty.

The invariant of the while-loop (lines 4 - 8) is that all states are sound, but setting a vertex to \texttt{DEAD} may lead to its predecessors to be incomplete; also, all vertices in $Q$ have been set to \texttt{DEAD} and all $v'.\degout$ are off by one where $v'$ are all vertices with a successor in $Q$,
\begin{dmath*}
sound(V)~\land~complete(V' \setminus Q.\pre)~\land~(\forall v \in Q : v.\status = {\DEAD})
~\land~(\forall v \in V : v.\degout = |\{u \in v.\post \mid u.\status = {\LIVE} \lor u \in Q\}|)
\end{dmath*}
where $Q.\pre = (\mathop{\cup} q \in Q : q.\pre)$. Since the while-loop terminates only when $Q = \emptyset$, it follows that the invariant of the outer for-loop is preserved.
We now argue that the while-loop preserves this invariant:
\begin{itemize}
\item $sound(V)$ is preserved as $v\in V$ is set to {{\DEAD}} only if $v.\degout = 0$, which implies that there cannot be {{\LIVE}} vertices in $v.\post$.
\item $complete(V' \setminus Q.\pre))$ is preserved as $v\in V'\setminus Q.\pre$ is completed vertices and $v$ is indeed set to {\DEAD} if $v.\degout = 0$, which implies that there cannot be {\LIVE} vertices in $v.\post$.
\item $\forall v \in Q : v.\status = {\DEAD}$ is preserved as $v$ is added to $Q$ only after $v$ is set to \DEAD.
\item $\forall v \in V : v.\degout = |\{u \in v.\post \mid u.\status = {\LIVE} \lor u \in Q\}|$ is preserved as $v.\degout$ is initialized as the number of available out-going edges and decremented only when removed successors propagated.
\end{itemize}

At termination of outer for-loop (lines 2 - 8 ), we get $Q=\emptyset$ and $V'=V$. The postcondition of outer for-loop is $sound(V) \land complete(V)$.
\qed \end{proof}






\fi 

\paragraph{Complexities.} 
\begin{theorem}
The worst-case time complexity of Algorithm \ref{Alg:seq-ac4trim} is $\mathcal{O}(n+m)$.
\begin{proof} \rm
The out-degree counter $v.\degout$ for all vertices can be initially calculated within $\mathcal{O}(n+m)$ time (line 1) as each edge is traversed once. Each vertex $v\in V$ can be removed and then added into the waiting set $Q$ at most once (lines 10 and 11); each reversed edge in $v(G^T).\post$ is traversed at most once (lines 6 - 8). In this case, in lines 2 - 8, we get a running time of $\mathcal{O}(n+m)$. Therefore, the total worst-case running time is $\mathcal O(n+m)$. 
\qed \end{proof} 
\end{theorem}

\begin{theorem}
The space complexity of Algorithm \ref{Alg:seq-ac4trim} is $\mathcal{O}(n+m)$.
\begin{proof} \rm
In line 7, the transposed graph $G^T = (V, E^T)$ is used. In this case, in order to generate $G^T$, the whole graph $G=(V,E)$ must be stored in memory, which requires $\mathcal{O}(n+m)$ space. For all vertices $v\in V$, storing $v.\degout$ and $v.\status$ uses $\mathcal O(n)$ space. Therefore, the total used space is $\mathcal O(n+m)$.   
\qed \end{proof} 
\end{theorem}

\subsection{The Parallel AC-4-Based Algorithm}
\label{sec:parallel-ac-4-based}
Algorithm \ref{Alg:par-ac4trim} shows the detailed steps of the parallel AC-4-based algorithm. All vertices in $V$ are partitioned into $V_1\dots V_{\mathcal{P}}$ (line 2) so that $\mathcal P$ workers can execute the procedure $\texttt{Trim}_p(V_p)$ in parallel (line 3).
Compared with Algorithm \ref{Alg:seq-ac4trim}, there are three refinements. 
First, each worker $p \in [1...\mathcal{P}]$ has its private waiting set $Q_p$ for propagation (line 6) so that the operations on $Q_p$ do not require to be synchronized.
Secondly, the out-degree counter $\degout$ has to be updated by the atomic primitive fetch\&add \texttt{FAA} since multiple workers may decrease such a counter (line 11). 
Thirdly, it is possible that $v.\degout = 0$ (in line 13) is detected by multiple workers; we use the atomic primitive \texttt{CAS} to set the $v.\status$ from $\LIVE$ to $\DEAD$ (line 13) and return $\TRUE$ if successful, which ensures that $v$ is added into a single one waiting set $Q_p$ (line 14).

\begin{algorithm}[htb]
\caption{Parallel AC-4-based Graph Trimming}
\label{Alg:par-ac4trim}
\DontPrintSemicolon
\SetAlgoLined
\SetAlgoNoEnd
\SetKwFunction{ContractSCC}{ContractSCC}%
\SetKwFunction{Contract}{Contract}
\SetKwFunction{Core}{Core}
\SetKwFunction{First}{FirstPost}
\SetKwFunction{Indegree}{InDegree}
\SetKwFunction{Outdegree}{OutDegree}
\SetKwFunction{CAS}{CAS}
\SetKwFunction{FAA}{FAA}
\SetKwFunction{Next}{DoDegree$_p$}
\SetKwFunction{Trim}{Trim$_p$}
\SetKwFunction{TrimOne}{Trim$_1$}
\SetKwFunction{TrimP}{Trim$_{\mathcal{P}}$}
\SetKw{Continue}{continue}
\SetKwData{Live}{\small LIVE}
\SetKwData{Dead}{\small DEAD}
\SetKwData{Gray}{gray}
\SetKwData{Unseen}{unseen}
\SetKwData{Instack}{instack}
\SetKwData{Complete}{complete}
\SetKw{With}{with}
\SetKwInOut{Input}{input}\SetKwInOut{Output}{output}
\Input{Graph $G=(V, E)$ and its transposed graph $G^T=(V, E^T)$}
\Output{Trimmed graph $G$}
\lFor{$v\in V$}{$v.\status, v.\degout := {\LIVE}, |v.\post|$}
partition $V$ into $V_1, \dots, V_{\mathcal P}$\;
\TrimOne{$V_1$} $\parallel \dots \parallel$ \TrimP{$V_{\mathcal P}$}

\medskip

\SetKwProg{myproc}{procedure}{ \Trim{$V_p$}}{}
\myproc{}{
    \For(){$v \in V_p$ \With $v.\status = \LIVE$}{
        $Q_p := \emptyset$; \Next{$v$}\;
        \While(){$Q_p \neq \emptyset$}{
            remove a vertex $w$ from $Q_p$\;
            \For{$v'\in w(G^T).\post$}{
                \FAA{$v'.\degout, -1$}\;
                \Next{$v', Q_p$}
            }
        }
    }
}

\medskip
\SetKwProg{myproc}{procedure}{ \Next{$v, Q_p$}}{}
\myproc{}{
    \If{$v.\degout = 0~\land~$\CAS{$v.\status$, \rm {\LIVE}, \rm {\DEAD}}}{
        $Q_p := Q_p \cup \{v\}$
    }
   
}

\end{algorithm}

\paragraph{Correctness.}
We show the soundness and completeness together.

\begin{theorem}[Soundness and Completeness]
For any $G=(V,E)$ Algorithm~\ref{Alg:par-ac4trim} terminates with $sound(V)$ and $complete(V)$.
\end{theorem}
\begin{proof}
The invariant of the while-loop (lines 4 - 8) in procedure $\texttt{Trim}_p(V_p)$ is the same as that in Algorithm \ref{Alg:seq-ac4trim} except that it adds one more conjunct. 
That is, a removed vertex can only be added into a single one $Q_p$ for propagation. 
\begin{dmath*}
sound(V_p) \land complete(V'_p \setminus Q_p.\pre) \land (\forall v \in Q_p : v.\status = {\DEAD}) 
\land (\forall v \in V : v.\degout = |\{u \in v.\post \mid u.\status = {\LIVE} \lor u \in \mathop{\cup} Q_{1..\mathcal P}\}|) 
\land (\forall i,j \in \{1 .. \mathcal P\}: i \neq j~\implies~Q_i \cap Q_j = \emptyset)
\end{dmath*}

We now argue that the while-loop preserves this invariant:
\begin{itemize}
\item $(\forall v \in V : v.\degout = |\{u \in v.\post \mid u.\status = {\LIVE} \lor u \in \mathop{\cup} Q_{1..\mathcal P}\}|)$ is preserved as  $v.\degout$ is off by one atomically when a worker is decreasing.
\item $(\forall i,j \in \{1..\mathcal P\}: i \neq j \land Q_i \cap Q_j = \emptyset)$ is preserved as $v.status$ is set from $\LIVE$ to $\DEAD$ by the atomic primitive $\texttt{CAS}$ and only when successful, $v$ is added to one $Q_p$.  
\end{itemize}

The postcondition of line 3 is then $sound(V_1)\land complete(V_1) \dots sound(V_{\mathcal{P}})\land complete(V_{\mathcal{P}})$, which is equivalent to $sound(V)\land complete(V)$. 
\qed \end{proof}

\paragraph{Complexities.} 
\begin{theorem}
Algorithm \ref{Alg:par-ac4trim} requires $\mathcal O(n+m)$ expected work, $\mathcal O (|Q_p| \Degin \Degout)$ depth, and thus $\mathcal O ((n+m)/\mathcal P+ |Q_p| \Degin \Degout)$ time complexity.
\begin{proof} \rm
This algorithm has the same framework as Algorithm \ref{Alg:seq-ac4trim}, so the total expected work equals the running time of Algorithm \ref{Alg:seq-ac4trim}, that is $\mathcal O(n+m)$.
The initial for-loop (lines 1) can easily run in parallel within expected depth $\mathcal O(\Degout)$.

We analyze the working depth for the procedure \texttt{Trim$_p$}. For each round of the outer while-loop (lines 7 - 11), it runs with depth $|Q_p|$ which is the upper-bound size of waiting sets among $\mathcal{P}$ workers. As $Q_p$ is private for worker $p$ without synchronization, it is possible that $|Q_p| \ge \alpha$. The most inner for-loop (line 9) runs sequentially with depth $\mathcal O(\Degin)$, and the out-degree counters have to concurrently update with depth $\Degout$. 
Therefore, the total working depth is $\mathcal O(\alpha \Degin \Degout)$ and thus the worst-case time complexity is $\mathcal O ((n+m)/\mathcal P+\alpha \Degin \Degout)$.
\qed \end{proof}
\end{theorem}

\begin{theorem}
The space complexity of Algorithm \ref{Alg:par-ac4trim} is $\mathcal O(n + m)$.
\begin{proof} \rm
By using the atomic primitive \texttt{CAS} in line 12, each vertex $v\in V$ may be removed at most once and then put into at most single one waiting set $Q_p$, so all waiting set for $\mathcal P$ workers require $\mathcal{O}(n)$ space. Similar to Algorithm \ref{Alg:seq-ac4trim}, storing $\degout$ and $\status$ requires $\mathcal O(n)$ space and the reverse edges require $\mathcal O (n+m)$ space (line 8). Therefore, the total used space is $\mathcal O(n + m)$.
\qed \end{proof}
\end{theorem}

\section{AC-6-Based Graph Trimming}
As mentioned, AC-4 has better worst-case time complexity than AC-3, but AC-4 always has a worse average running time than AC-3. Additionally, AC-4 does not have the on-the-fly property.
AC-6 improves AC-4 by only recording one support for each value since one support is enough to guarantee that a value is viable. In this case, compared with AC-4, AC-6 performs better in many applications, requires less space usage, and has the on-the-fly property. 

To the best of our knowledge, we are the first to introduce AC-6 to graph trimming and call it the AC-6-based algorithm. The idea is novel: 1) each vertex~$v$ maintains a set of vertices $v.S$ that choose $v$ as an available outgoing edge; 2) when removing $v$ as it has no outgoing edges, each vertex $w\in v.S$ has to find another available outgoing edge to replace $v$; otherwise, $w$ has to be removed; 3) this process repeats until no vertices can be removed.
In this section, we propose new sequential and parallel AC-6-based algorithms for graph trimming, which is the main contribution of this work.

\subsection{The Sequential AC-6-Based Algorithm}

Analogous to AC-6, the AC-6-based trimming algorithm is based on the concept of \emph{support}. That is, for each vertex $v$ in a given directed graph $G$, the support of $v$ is one of $v$'s available outgoing edges and $v$ cannot be removed if $v$'s support exists. One auxiliary data structure, the supporting set, is needed to store all the supports for propagation, which is formally defined below. 

\begin{definition}[Supporting Set] \label{def:supporing}
Given a directed graph $G=(V, E)$, for a vertex $v\in V$, the supporting set $v.S$ of $v$ is the set of predecessors that choose $v$ as their single one support: $(\forall v\in V: v.S\subseteq\{u\in v.\pre \mid u.\status = \LIVE\})~\land~(\forall v\in V: v.S \neq \emptyset \implies v.status = \LIVE)~\land~(\forall u, v\in V: u\neq v \implies u.S \cap v.S = \emptyset)$.
\end{definition}

In other words, $v.S$ records all \LIVE~vertices that have $v$ as their support. Absolutely, $v$ must be \LIVE~if the vertices in $v.S$ choose $v$ as a support as an existing support has to be an available outgoing edge; 
a vertex can be added into at most one supporting set as each vertex only needs to maintain single one support.

\begin{figure}[htb]
\centering
\includegraphics[scale=0.8]{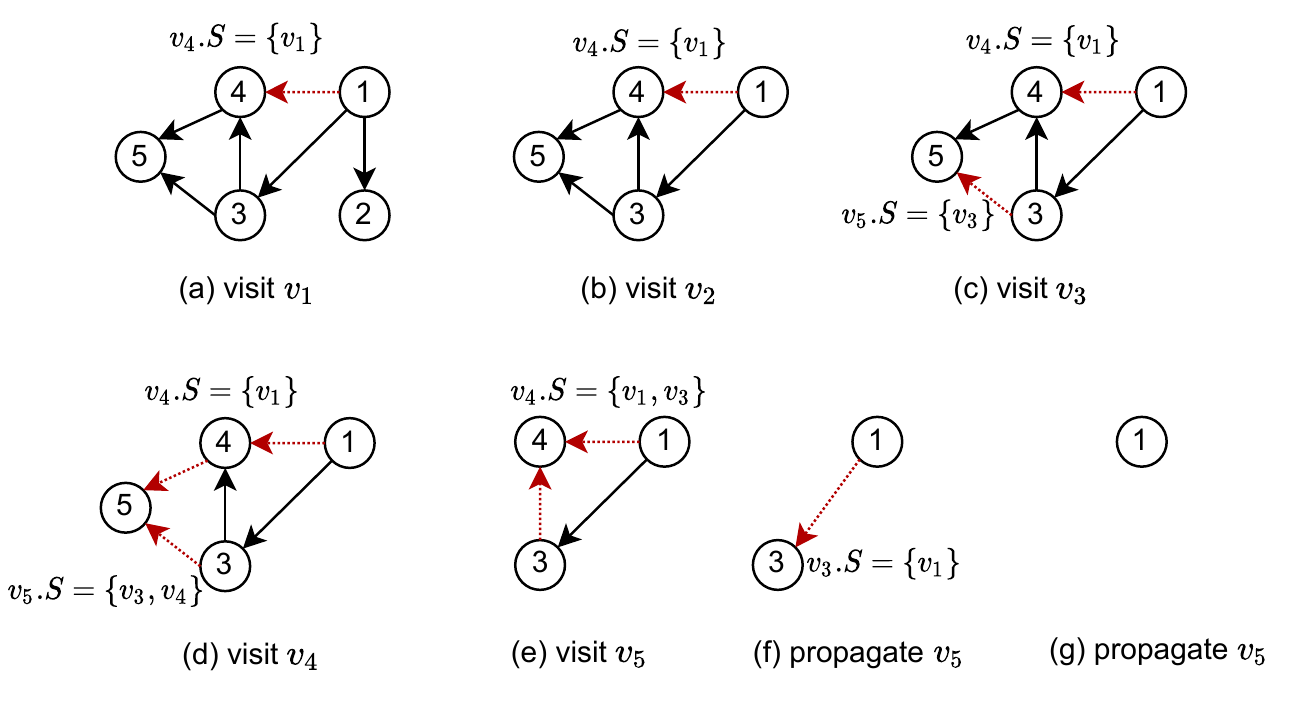}
\caption{Steps of the sequential AC-6-based trimming algorithm based on part of the graph in Figure~\ref{Figure:example-graph}.}
\label{Fig:fast-trim}
\end{figure}


Figure~\ref{Fig:fast-trim} illustrates the AC-6-based algorithm based on the part of the example graph in Figure~\ref{Figure:example-graph}. The dashed red arrows are the edges visited to find available supports and then added to the corresponding supporting set $v.S$. 
Each vertex is successively visited from $v_1$ to $v_5$ as shown in Figure~\ref{Fig:fast-trim}~(a) to (e).
In Figure~\ref{Fig:fast-trim}~(a), $v_1$ is visited and its first support $v_4$ is found with vertex $v_1$ added into $v_4.S$. 
In Figure~\ref{Fig:fast-trim}~(b), $v_2$ is removed since $v_2$ has no outgoing edges; no propagation happens as $v_2.S$ is empty. 
In Figure~\ref{Fig:fast-trim}~(c), $v_3$ is visited and its first support $v_5$ is found with $v_3$ added into $v_5.S$. 
In Figure~\ref{Fig:fast-trim}~(d), $v_4$ is visited and its first support $v_5$ is found with $v_3$ added into $v_4.S$. 
In Figure~\ref{Fig:fast-trim}~(e), $v_5$ is removed as it cannot find any support; since $v_5.S$ includes $v_3$ and $v_4$ the propagation happens as follow: $v_3$ finds a next available support $v_4$ with $v_3$ added into $v_4.S$, and the $v_4$ is failed to find a next available support so that $v_4$ should be removed in the next step.
In Figure~\ref{Fig:fast-trim}~(f), $v_4$ is removed; since the supporting set $v_4.S$ includes vertices $v_1$ and $v_3$ the propagation happens as follow: $v_1$ finds a next available support, $v_3$, which is added into $v_4.S$, and $v_3$ fails to find a next available support so that $v_3$ should be removed in the next step. 
In Figure~\ref{Fig:fast-trim}~(g), the vertex $v_3$ is removed and $v_1\in v_3.S$ should be further propagated.
Finally, $v_1$ should also be removed as it has no outgoing edges. 
As we can see, AC-6-based trimming can remove some of the vertices without propagation, e.g. $v_2$.

\begin{algorithm}[htb]
\caption{Sequential AC-6-based Graph Trimming}
\label{Alg:seq-ac-6-trim}
\DontPrintSemicolon
\SetAlgoLined
\SetAlgoNoEnd
\SetKwFunction{ContractSCC}{ContractSCC}%
\SetKwFunction{Contract}{Contract}
\SetKwFunction{Next}{DoPost}
\SetKwFunction{Core}{Core}
\SetKwFunction{First}{FirstPost}
\SetKw{Continue}{continue}
\SetKw{With}{with}
\SetKwData{Live}{\small LIVE}
\SetKwData{Dead}{\small DEAD}
\SetKwData{Gray}{gray}
\SetKwData{Unseen}{unseen}
\SetKwData{Instack}{instack}
\SetKwData{Complete}{complete}
\SetKwInOut{Input}{input}\SetKwInOut{Output}{output}
\Input{Graph $G=(V, E)$}
\Output{Trimmed graph $G$}

\lFor{$v\in V$}{$v.\status, v.S := {\Live}, \emptyset$}

\For{$v \in V$}{
    $Q := \emptyset$; \Next{$v$} 
    
    \While{$Q\neq \emptyset$}{
        remove a vertex $w$ from $Q$\; 
        \For{$v' \in w.S$}{
            $w.S := w.S \setminus \{v'\}$\;
            \Next{$v'$}
        }
    }

}
\medskip
\SetKwProg{myproc}{procedure}{ \Next{$v$}}{}
\myproc{}{
    \For{$ w \in v.\post$ \With $w.\status = \Live$}{
            $w.S := w.S \cup \{v\}$; $v.post:=v.post\setminus \{w\}$; \Return
    }
    {$v.\status := {\Dead}$; $Q:= Q \cup \{v\}$}
    
}
\end{algorithm}

Algorithm~\ref{Alg:seq-ac-6-trim} shows the detailed steps of the sequential AC-6-based algorithm. 
For each vertex $v$ in the graph, a supporting set $v.S$ is required for recording the vertices that choose $v$ as an available support. We first consider the procedure $\texttt{DoPost}(v)$ (lines 9 - 12). 
If $v$ successfully finds a live successor $w$, then $w$ is added to $v.S$ and the procedure finishes (lines 10 and 11). Otherwise, $v$ has to be removed from the graph as $v$ has no available outgoing edges, and $v$ is set to \texttt{DEAD} and then put into the waiting set~$Q$ (line 12). Note that, the visited vertex $w$ is removed from $v.post$ to avoid redundant checking (line 11), which can ensure that each edge is visited at most once.
Now we explain the main algorithm (lines 1 - 8). Initially, all vertices are \LIVE~and their supporting sets are empty (line 1).
For each vertex $v\in V$, the support $v.s$ is checked by the procedure $\texttt{DoPost}(v)$ (line 3) and the removed vertices are added into $Q$ for propagation (lines 4 - 8). 
That is, a vertex $w\in Q$ is removed from $Q$ (line 5) and for all the vertices in $w.S$ are checked by the procedure $\texttt{DoPost}(v')$ (lines 6 - 8). This propagation is repeated until $Q$ is empty (line 4), as vertices may be removed and added into $Q$ by the procedure $\texttt{DoPost}(v')$ (line 8).

\paragraph{Correctness.}
We show the soundness and completeness together.
\begin{theorem} [Soundness and Completeness]
For any $G=(V,E)$ Algorithm~\ref{Alg:seq-ac-6-trim} terminates with $sound(V)$ and $complete(V)$.
\end{theorem}
\begin{proof}
Let $V'$ be the set of vertices visited by the outer for-loop (lines 2 - 8). The invariant of the outer for-loop is that all vertices are sound, all visited vertices are complete, all visited \texttt{LIVE} vertices must have a support, and that for each vertex $v\in V$ the supporting set $v.S$ includes all visited vertices that choose $v$ as their single one support:
\begin{dmath*} 
sound(V)~\land~complete(V')
~\land~(\forall v \in V': v.status = \LIVE \implies v\in S)
~\land~(\forall v \in V : v.S \subseteq \{u \in v.\pre \mid u.\status = \LIVE \land u\in V'\})
~\land~(\forall v\in V: v.S \neq \emptyset \implies v.status = \LIVE)
~\land~(\forall u, v\in V: u\neq v\implies u.S\cap v.S = \emptyset)
\end{dmath*}
where $S = (\mathop{\cup} v\in V: v.S)$.
The invariant holds initially as setting all vertices to $\LIVE$ and $V'$ is empty. 

The invariant of the while-loop (lines 4 - 8) is that all states are sound but setting a vertex to $\DEAD$ may lead to its predecessors to be incomplete; also, all vertices $w\in Q$ are set to $\DEAD$ and all vertices in $w.S$ have to update their support.
\begin{dmath*}
sound(V)~\land~complete(V'\setminus Q.S)~\land~(\forall v\in Q: v.status = {\DEAD}) 
~\land~(\forall v \in V': v.status = \LIVE \implies v\in S)
~\land~(\forall v \in V : v.S \subseteq \{u \in v.\pre \mid u.\status = \LIVE \land u\in V'\})
~\land~(\forall v\in V: v.S \neq \emptyset \implies v.status = \LIVE \lor v \in Q)
~\land~(\forall u, v\in V: u\neq v\implies u.S\cap v.S = \emptyset) 
\end{dmath*}
where $S = (\mathop{\cup} v\in V: v.S)$ and $Q.S = (\mathop{\cup} q\in Q: q.S)$. Since the while-loop terminates only when $Q=\emptyset$, it follows that the invariant of the outer for-loop is preserved.

We now argue that the while-loop preserves this invariant: 
\begin{itemize}
\item $sound(v)$ is preserved as $v\in V$ is set to \texttt{DEAD} if $w$ cannot find a support in $v.post$, which implies that there cannot be \texttt{LIVE} vertices in $v.post$.
\item $complete(V'\setminus Q.S)$ is preserved as $v\in V'\setminus Q.S$ is indeed set to \texttt{DEAD} if the support of $v$ not exists, which implies that there cannot be \texttt{LIVE} vertices in $v'.post$.
\item $\forall v\in Q: v.status = {\DEAD}$ is preserved as $v$ is added to $Q$ only after $v$ is set to \texttt{DEAD}.
\item $\forall v \in V': v.status = \LIVE \implies v\in S$ is preserved as $v$ has to find a support after being visited if $v$ is \LIVE. 
\item $\forall v \in V : v.S \subseteq \{u \in v.\pre \mid u.\status = \LIVE \land u\in V'\}$ is preserved as the visited vertices $u\in V'$ is \LIVE~when choosing $v$ as a support.
\item $\forall v\in V: v.S \neq \emptyset \implies v.status = \LIVE \lor \land v\in Q$ is preserved as new vertices can be added into $v.S$ if $v$ is $\LIVE$, and after setting $v$ to $\DEAD$ and adding into $Q$~all vertices in $v.S$ will find next available support.
\item $\forall u, v\in V: u\neq v\implies u.S\cap v.S = \emptyset$ is preserved as each vertex maintains at most single one support.
\end{itemize}

At termination of the outer for-loop (lines 2 - 8), we get $Q=\emptyset$ and $V'=V$. The postcondition of the outer for-loop is $sound(V)\land complete(V)$.
\qed  \end{proof}

\paragraph{Complexities.} 
\begin{theorem}
The time complexity of the Algorithm \ref{Alg:seq-ac-6-trim} is $\mathcal{O}(n+m)$.
\begin{proof} \rm

Each vertex $v\in V$ can be removed and then added into the waiting set $Q$ at most once.
In the procedure \texttt{DoPost($v$)}, each outgoing edge of $v$ is traversed at most once to find the support (lines 10 and 11) as the visited vertex $w$ is removed from $v.post$ to avoid repetitive visiting. In this case, 
The most inner for-loop (lines 6 - 8) calls procedure \texttt{DoPost($v$)} to find a support for vertex $v$.
Therefore, with this assumption, the worst-case time complexity is~$\mathcal{O}(n+m)$. 
\qed \end{proof}
\end{theorem}

\begin{theorem}
\label{th:par-ac4-space}
The space complexity of the Algorithm \ref{Alg:seq-ac-6-trim} is $\mathcal{O}(n)$.
\begin{proof} \rm
The global waiting set $Q$ has a maximum size of $\mathcal{O}(n)$ as each vertex $v\in V$ can be set to \texttt{DEAD} and added into $Q$ at most once. The supporting sets have the total size at most $\mathcal{O}(n)$ as each vertex $v\in V$ has at most one support recorded in a corresponding supporting set. Obviously, $\status$ requires $\mathcal{O}(n)$ space. Therefore, the worst-case space complexity is $\mathcal{O}(n)$. 
\qed \end{proof}
\end{theorem}

\subsection{The Parallel AC-6-Based Algorithm}
Algorithm \ref{Alg:par-ac-6-trim} shows the detailed steps of the parallel AC-6-based trimming algorithm.
Compared with the sequential AC-6-based trimming in Algorithm \ref{Alg:seq-ac-6-trim}, there are two refinements.
First, each worker $p\in [1\dots \mathcal{P}]$ has its private waiting set $Q_p$ for propagation so that the synchronization on $Q_p$ is unnecessary (lines 6, 8, and~19). Secondly, the supporting set $w.S$ for each vertex $w\in V$ can concurrently have new vertices added by multiple workers synchronized by locking (lines 14 - 17). 
When adding vertices to $w.S$, vertex $w$ has to be \LIVE~(line 15) as no vertices can be added to $w.S$ after setting $w$ to \DEAD. 
When setting vertex $v$ to \DEAD, we have to lock $v$ to ensure that no other workers are adding vertices to $v.S$; otherwise, after $v$ is added into $Q$ and propagated (lines 19 and 8 - 11), other workers still have possibility to add vertices into $v.S$ which can never be propagated. 
In other words, we lock $v.S$ when setting $v$ from \LIVE~to \DEAD~to ensure that all vertices in $v.S$ are propagated together.
Note that, when removing vertices from $w.S$ (line 10), it is unnecessary to lock $w.S$ as currently $w$ is \texttt{DEAD} so that no workers can add vertices into $w.S$ and $w$ is only accessed by a single worker, $p$. 

We implement the lock by the \texttt{CAS} primitive with busy waiting (lines 20 and 21). Here, the busy waiting is suitable as there are at most two operations within locking (lines 15 and 16) so that the expected waiting time is really short.

\begin{algorithm}[htb]
\caption{Parallel AC-6-based Graph Trimming}
\label{Alg:par-ac-6-trim}
\DontPrintSemicolon
\SetAlgoLined
\SetAlgoNoEnd
\SetKw{Continue}{continue}
\SetKwFunction{ContractSCC}{ContractSCC}
\SetKwFunction{CAS}{CAS}
\SetKwFunction{Lock}{Lock}
\SetKwFunction{Unlock}{Unlock}
\SetKwFunction{Contract}{Contract}
\SetKwFunction{Next}{DoPost$_p$}
\SetKwFunction{Trim}{Trim$_p$}
\SetKwFunction{TrimOne}{Trim$_1$}
\SetKwFunction{TrimP}{Trim$_{\mathcal P}$}
\SetKwFunction{ListDelete}{Delete}
\SetKwFunction{First}{FirstPost}
\SetKwFunction{AddSupport}{AddSupport}
\SetKwFunction{Lock}{Lock}
\SetKwFunction{Unlock}{Unlock}
\SetKwFunction{CAS}{CAS}
\SetKw{Continue}{continue}
\SetKw{With}{with}
\SetKwData{Live}{\small LIVE}
\SetKwData{Dead}{\small DEAD}
\SetKwData{True}{\small TRUE}
\SetKwData{False}{\small FALSE}
\SetKwData{InQueue}{in-queue}
\SetKwData{Complete}{complete}
\SetKwData{Propagate}{propagate}
\SetKwData{Supporting}{supporting}
\SetKwData{Complete}{complete}
\SetKwInOut{Input}{input}\SetKwInOut{Output}{output}
\Input{$G=(V, E)$}
\Output{Trimmed graph $G$}
\lFor(){$v \in V$}{ $v.S, v.\status:=\emptyset, {\Live}$; \Unlock{$v$} }
partition $V$ into $V_1 , \dots, V_{\mathcal{P}}$\;
\TrimOne{$V_1$} $\parallel \dots \parallel$ \TrimP{$V_{\mathcal{P}}$}\;

\medskip
\SetKwProg{myproc}{procedure}{ \Trim{$V_p$}}{}
\myproc{}{
    \For{$v\in V_p$}{ 
        $Q_p = \emptyset$; \Next{$v, Q_p$}\;
        \While(){$Q_p \neq \emptyset$}{
            remove a vertex $w$ from  $Q_p$\;
            \For{$v' \in w.S$}
            { $w.S := w.S \setminus \{v'\}$\; \Next{$v', Q_p$}}
        }
    }
}

\medskip
\SetKwProg{myproc}{procedure}{ \Next{$v, Q_p$}}{}
\myproc{}{
    \For{$w \in v.\post$ \With $w.\status = \Live$}{
        \Lock{$w$}\;
        \If{$w.\status = \Live$}{
            $w.S := w.S \cup \{v\}$;
            \Unlock{$w$}; \Return
        }
        \Unlock{$w$}\;
        
    }

    \Lock{$v$}; $v.\status := \Dead$; \Unlock{$v$}\;
    $Q_p:= Q_p \cup \{v\}$
    
}

\medskip
\SetKwProg{myproc}{procedure}{ \Lock{$w$}}{}
\myproc{}{
    \lWhile(){$\neg$~\CAS{$w.\lock, \False, \True$}}{ skip
    }
}

\medskip
\SetKwProg{myproc}{procedure}{ \Unlock{$w$}}{}
\myproc{}{
    $w.\lock:=\False$\;
}

    

\end{algorithm}

\paragraph{Correctness.}

\begin{theorem}[Soundness and Completeness of Parallel AC-6-based Trimming]
For any $G=(V,E)$ Algorithm~\ref{Alg:par-ac-6-trim} terminates with $sound(V)$ and $complete(V)$.
\end{theorem}
\begin{proof}
The invariant of the while-loop (lines 4 - 8) in procedure $\texttt{Trim}_p(V_p)$ is the same as that in Algorithm \ref{Alg:seq-ac4trim} except that it adds one more conjunct. 
That is, a removed vertex can only be added into a single one $Q_p$ for propagation. 
\begin{dmath*}
sound(V_p)~\land~complete(V'_p\setminus Q_p.S)~\land~(\forall v\in Q_p: v.status = {\DEAD}) 
~\land~(\forall v \in V'_p: v.status = \LIVE \implies v\in S)
~\land~(\forall v \in V_p : v.S \subseteq \{u \in v.\pre \mid u.\status = \LIVE \land u\in V'\})
~\land~(\forall v\in V_p: v.S \neq \emptyset \implies v.status = \LIVE \lor v\in Q_p)
~\land~(\forall u, v\in V_p: u\neq v\implies u.S\cap v.S = \emptyset) 
~\land~(\forall i,j \in \{1..\mathcal P\}: i \neq j \implies Q_i \cap Q_j = \emptyset)
\end{dmath*}
where $S = (\mathop{\cup} v\in V: v.S)$, $Q_p.S = (\mathop{\cup} q\in Q_p: q.S)$, and $V' =\mathop{\cup} V'_{1..\mathcal{P}}$. In the algorithm, for each vertex $v$, multiple workers add new vertices into $v.S$ concurrently, and during this time $v$ cannot be set to \DEAD. 
We now argue that the while-loop preserves this invariant:
\begin{itemize}
\item $\forall v \in V_p : v.S \subseteq \{u \in v.\pre \mid u.\status = \LIVE\land u\in V'\}$ is preserved as $v$ is locked when a worker is adding new vertices to $v.S$.
\item $\forall v\in V_p: v.S \neq \emptyset \implies v.status = \LIVE \lor (v.\status = \DEAD \land v\in Q_p)$ is preserved as 1) $v$ is locked when the worker $p$ setting $v$ to $\DEAD$ to ensure that after setting $v$ to \DEAD~no vertices can be added into $v.S$ by other workers, and 2) only the current worker $p$ can set $v$ to $\DEAD$ and add $v$ to the private set $Q_p$.
\item $\forall i,j \in \{1..\mathcal P\}: i \neq j~\land~Q_i \cap Q_j = \emptyset$ is preserved as only single one worker $p$ can add $v$ to $Q_p$ after setting $v$ to $\DEAD$.  
\end{itemize}

At the termination of outer for-loop (lines 2 - 8), we get $Q_p = \emptyset$ and $V'_p=V_p$.  
The postcondition of line 3 is then $sound(V_1)\land complete(V_1)\land \dots \land sound(V_{\mathcal{P}})\land complete(V_{\mathcal{P}})$, which is equivalent to $sound(V)\land complete(V)$. 
\qed \end{proof}

\paragraph{Complexities.}
\begin{theorem}
The Algorithm \ref{Alg:par-ac-6-trim} requires $\mathcal O(n+m)$ expected work, $\mathcal O (|Q_p| \Degin^2)$ depth, and $\mathcal O ((n+m)/\mathcal P+ |Q_p| \Degin^2)$ time complexity.
\begin{proof} \rm
This algorithm has the same framework as Algorithm \ref{Alg:seq-ac-6-trim}, so the total expected work equals the running time of Algorithm \ref{Alg:seq-ac-6-trim}, that is $\mathcal O(n+m)$.
The initial for-loop (lines 1) can run in parallel within expected depth $\mathcal O(1)$.

We analyze the working depth for procedure \texttt{Trim$_p$}. For each round of the outer while-loop (lines 7 - 11), it runs with depth $|Q_p|$ which is the upper-bound size of waiting sets among $\mathcal{P}$ workers. As $Q_p$ is private for a worker $p$ without synchronization, it is possible that $|Q_p| \ge \alpha$. The most-inner for-loop (line 9) runs sequentially with depth $\Degin$ as $\Degin$ is the upper-bound size for a supporting set. The supporting sets concurrently add new vertices with depth $\Degin$ (line 16).
The locking operation (lines 14 and 18) needs to busy-check by the \texttt{CAS} primitive only a few times with high probability as there are at most two operations within the lock.
Therefore, the total working depth is $\mathcal O(|Q|\Degin^2)$ with high probability and the worst-case running time is $\mathcal O((n+m)/\mathcal P+|Q|\Degin^2)$ with high probability. 
\qed \end{proof}
\end{theorem}



\begin{theorem}
The space complexity of Algorithm \ref{Alg:par-ac-6-trim} is $\mathcal O(n)$, which equals to its sequential version Algorithm \ref{Alg:seq-ac-6-trim}.
\begin{proof} \rm
Each vertex $v\in V$ may be removed at most once and then put into at most one waiting set $Q_p$, so all waiting sets require $\mathcal{O}(n)$ space. Each vertex has at most one support which is stored into the corresponding supporting set and thus the total size of all supporting sets is $\mathcal O (n)$. The status for each vertex $v \in V$ has the size of $\mathcal O(n)$. Therefore, the total space complexity is $\mathcal O(n)$.
\qed \end{proof}
\end{theorem}

\section{Related Work}

\subsection{Parallel DFS-based SCC Decomposition}

In Section 1, several methods for BFS-based SCC decomposition were introduced. Although DFS is inherently sequential~\cite{Reif1985}, there is a lot of work based on Tarjan's algorithm.
In~\cite{Lowe2016}, Lowe proposed a synchronized Tarjan’s algorithm, that is, multiple instances of Tarjan's algorithm run without overlapping stacks. To do this, a worker is suspended on a vertex which is located in another worker’s stack and then both workers' stacks can be merged if necessary. 
The drawback is that this stack merging leads to a worst-case quadratic time complexity of $\mathcal{O}(n^2)$. Lowe’s experiments show decent speedups on model checking graphs with trivial SCCs, but not for graphs with large SCCs. 
In~\cite{Renault2015}, Renault et al. present a novel algorithm without sacrificing the linear time complexity, $\mathcal{O}(n+m)$, and the on-the-fly property. Multiple instances of Tarjan’s algorithm run and communicate completely explored SCCs via a shared union-find structure. Bolomen et al.~\cite{bloemen2016multi,Bloemen2015} proposed an improved UFSCC algorithm which communicates partially found SCCs by using a modified union-find data structure. In their experiments, UFSCC shows a significant speedup compared to Renault's algorithm~\cite{Renault2015} on implicit model checking inputs and synthetic graphs.  
One notable property of these algorithms is that they can run on-the-fly on implicit graphs. 

However, above DFS-based SCC algorithms do not utilize graph trimming techniques to remove trivial SCCs. A possible reason is that the traditional graph trimming technique has quadratic worst-case time complexity and, more importantly, it is hard to run on-the-fly. The proposed parallel AC-6-based graph trimming algorithm has linear running time and has the on-the-fly property, so it can be used to quickly trimming a high ratio of \mbox{size-$1$} SCCs \del{reduce the search space}for above DFS-based SCC algorithms. 


\subsection{Graph Trimming}
Generally, the term ``graph trimming” is widely used in graph algorithms to minimizing the search space and the trimming rules may be different for different problems. 
For example, in \cite{heule2019trimming}, ``graph trimming" computes a smaller and smaller unsatisfiable core for a propositional formula; 
in \cite{gao2015scalpel}, ``graph trimming" is used to minimize the number of vertices as monitors to identify all interesting links; 
in \cite{erlebach2010trimming}, given a graph in which each vertex has a nonnegative weight, ``graph trimming" deletes vertices with a small total weight such that the remaining graph does not contain any long simple paths.
Note that, in the current work, given a directed graph, the terminology ``graph trimming” is specifically confined as each vertex has at least one outgoing edge.

Another related term is ``graph pruning".
For example, in \cite{harabor2011online}, given a geographical graph, ``graph pruning" can dynamically jump over some searching branches by some simple rules for finding the path between two nodes.

%% file: body-experiment.tex

\section{Implementation}
All graph trimming algorithms are implemented in C++ with OpenMP as treading library. In this section, we explore the details of implementation, especially the parallelism. 
\paragraph{Graph Storage.}
All tested graphs are stored in the \emph{compressed sparse row} CSR format \cite{Hong2012} for efficient traversals. 
With this format, all edges are linear in the memory. In other words, for each vertex the edges can be traversed in sequential order. 
In the current work, our experiments focus on graphs with edges linearly stored in memory.

\paragraph{Parallelism.}
OpenMP (Open Multi-Processing)~\cite{dagum1998openmp} is an application programming interface (API) that supports multi-platform shared-memory multiprocessing programming in C, C++, and Fortran, on many platforms, instruction-set architectures and operating systems. In this paper, OpenMP (version 4.5) is used as the threading library to implement the parallel algorithms. The task-level parallelism is implemented by using the clause ``\texttt{\#pragma omp parallel for}'' (C++ code). 
Given a input graph, this implementation statically assigns the same number of vertices to each worker $p$.
For data-level parallelism, however, it is critical to handle a potential workload imbalance problem. 
Note that real-world graphs can be highly irregular because of their scale-free property, e.g., a few vertices can have a huge number of successors while many vertices have only several successors. 
Therefore, statically assigning the same number of vertices to each worker naturally induces workload imbalance since the work of each vertex involves immediate propagation. 
There is a better strategy. All of the vertices in the graph can be dynamically assigned to each worker $p$ by the clause ``\texttt{\#pragma omp for schedule(dynamic, $s$)}''. That means each worker executes a chunk of iterations with size $s$ and then requests another chunk until no chunks remain to distribute. 
If one of the workers finishes processing a chunk of vertices early, it applies to the next chunk of vertices at once without waiting for other workers. In this way, we realize a relatively balanced load for each worker without difficulties.
Note that the chunk size cannot be either much large or small; the too large chunk size may cause work-load imbalance for multiple workers; the too small chunk size may cause much running time spent on scheduling.  

For simplicity, our parallel trimming algorithms sacrifice some parallelism. That is, the most inner \mbox{for-loop} can run in parallel (lines 12 - 13 in Algorithm \ref{Alg:AC-3}, lines 9 - 11 in Algorithm \ref{Alg:par-ac4trim}, and lines 9 -11 in Algorithm \ref{Alg:par-ac-6-trim}); the private waiting set $Q_p$ for a worker $p$ can be balanced in Algorithm \ref{Alg:par-ac4trim} and Algorithm \ref{Alg:par-ac-6-trim}) so that $Q_p$ has at most $\alpha$ vertices. As shown in Table \ref{Table:trim-2},  the working depth can be improved if we achieve full parallelism. However, the scheduler will be challenged to parallel inside each worker $p$ efficiently. One solution is to maintain a \emph{frontier} (subset) of all affected vertices, and in each step all vertices in a frontier can be processed in parallel \cite{defo2019parallel}.

\begin{table*}[!htb]
\centering
\small
\begin{tabular}{c|c c c}
 \toprule
        & \multicolumn{3}{c}{Worst-Case ($\mathcal{O}$)} \\
 Trimming   &    Work & Depth & Space     \\     
 \midrule
AC-3-based  & $\alpha (n+m)$ & $\alpha $        &$n$\\ 
AC-4-based  & $n+m$         & $\alpha \Degout$  & $n+m$\\  
AC-6-based  & $n+m$         & $\alpha \Degin$   & $n$  \\ 
 \bottomrule
 
\end{tabular}
\caption{The worst-case work, depth, time, and space complexities of full parallelized graph trimming algorithms.}
\label{Table:trim-2}
\end{table*}

In practice, our algorithms can be highly parallelized. There are two reasons. First, most real graphs always have millions of edges, and $|Q|$, $\Degin$, and $\Degout$ are relatively much smaller than $n+m$. Secondly, multi-core machines always have a limited number of workers, e.g. $\mathcal P=32$. Therefore, our trimming algorithms can achieve a load balance among multiple workers with high probability.

\paragraph{Traverse Edges.}
Since the edges are linearly stored in memory, we can optimize the implementation of trimming algorithms. In the procedure \texttt{ZeroOutDegree} of Algorithm \ref{Alg:TraditionalTrim}, for vertex $v$ only the first \LIVE~edge needs to be found. In this case, each vertex can maintain an index $edge\_index$ to record the position of visited edges. In the next round, we can ``jump'' over the edges that have already visited. By doing this, we can reduce the number of traversed edges to a certain degree.
Similarly, we can apply this strategy to the procedure \texttt{DoPost} of Algorithm \ref{Alg:seq-ac-6-trim} and Algorithm \ref{Alg:par-ac-6-trim}; by doing this, each edge can be traversed at most once. 

\paragraph{Cache-Friendliness.}
For multi-core architectures, contiguous memory accessing is much faster than random memory accessing because of the possibility of pre-fetching by L1, L2, and L3 caches. Of course, accessing cache is faster than accessing the memory by an order of magnitude. For explicit graphs stored in memory, the cache can affect the running time by an order of magnitude. A \emph{cache-friendly} program has a large portion of contiguous memory accessing that can fully utilize the cache to obtain speedup. In contrast, a \emph{cache-unfriendly} program has a large portion of random memory accessing that can not efficiently utilize the cache. 



Since all edges are stored in contiguous memory as CSR format for a tested graph, we compare the cache property of three different graph trimming methods together as follow: 
\begin{itemize}
    \item AC-3-based Graph Trimming is cache-friendly as all edges are stored in an array and can be traversed sequentially with a high cache hit rate.
    \item AC-4-based Graph Trimming is less cache-friendly as each vertex $v$ are traversed almost randomly, but $v$'s edges are traversed sequentially with a medium cache hit rate.  
    \item AC-6-based Graph Trimming is least cache-friendly as for each vertex $v$, both $v$ and $v$'s edges are traversed almost randomly with low cache hit rate.  
\end{itemize}

\paragraph{Memory Usage.}
We compare the practical memory usage in Table \ref{tab:memory}. Assume that storing a vertex or an integer takes $H$ bits. 
All three algorithms require $1$ bit for the status of each vertex, in total $n$ bits.  
For both \emph{AC4Trim} and \emph{AC6Trim}, there are $\mathcal{P}$ waiting sets $Q_{1} \dots Q_{\mathcal{P}}$, in total $nH$ bits, since each vertex can be put into $Q_p$ at most once. 
For \emph{AC4Trim}, a reversed graph has to be loaded into memory, in total $(n+m)H$ bits; each vertex maintains an out-degree counter $\degout$, in total $nH$ bits.    
For \emph{AC6Trim}, each vertex has a supporting set $S$, in total $nH$ bits, since each vertex can be put into a set $S$ at most once.
For \emph{AC3Trim} and \emph{AC6Trim}, each vertex maintains an index $edge\_{index}$ to ``jump'' over the visited edges, in total $nH$ bits. 
\begin{table}[!htb]
\centering
\begin{tabular}{ l | l | l}
 \emph{AC3-based}        &  \emph{AC4-based}       & \emph{AC6-based}   \\ 
   \midrule
 bit[$n$]: $\forall v.\status$  & bit[$n$]: $\forall v.\status$         & bit[$n$]: $\forall v.\status$ \\ 
 bit[$nH$]:$\forall v.edge\_index$& bit[$nH$]: $\forall v.\degout$         & bit[$n$]: $\forall v.\lock$   \\
                               & bit[$nH$]: $Q_1\dots Q_{\mathcal{P}}$  & bit[$nH$]:$\forall v.edge\_index$ \\ 
                               & bit[$(n+m)H$]: $G^T$                   & bit[$nH$]: $\forall v.S$ \\
                               &                                        & bit[$nH$]: $Q_1\dots Q_{\mathcal{P}}$\\ 
\end{tabular}
\caption{Compare the memory usage for \emph{AC3Trim}, \emph{AC4Trim} and \emph{AC6Trim}, where storing a vertex takes $H$ bits.}
\label{tab:memory}
\end{table}

\section{Experiments}
In this section, we evaluate three different parallel algorithms for graph trimming: 
\begin{itemize}
    \item the AC-3-based trimming algorithm~\cite{hong2013fast,iSpanJi} (\emph{AC3Trim} for short),
    \item the AC-4-based trimming algorithm~\cite{McLendon2005} (\emph{AC4Trim} for short),
    \item the AC-6-based trimming algorithm (\emph{AC6Trim} for short). 
\end{itemize}
The experiments are performed on a server with an AMD CPU (16 cores, 32 hyper-threads, 32 MB of last-level cache) and 96 GB main memory. The server runs the Ubuntu Linux (18.04) operating system. All tested algorithms are implemented in C++ and compiled with g++ version 7.3.0 with the -O3 option
\footnote{All our implementations, benchmarks, and results are available at \url{https://github.com/Itisben/graph-trimming.git}}.
OpenMP~\cite{dagum1998openmp} version 4.5 is used as the threading library. We perform every experiment at least 50 times (at least 10 times for time-consuming experiments) and calculate their means with 95\% confidence intervals.

We first give in total 15 real and synthetic benchmark graphs. 
Before the evaluation, we discuss the workload balance.
Then, over these tested graphs, we evaluate the number of traversed edges and then compare the real running times by varying the workers from 1 to 32. 
We also evaluate the stability and the scalability by using 16 workers.

\subsection{Graph Benchmarks}
We evaluate the performance of our method on a variety of model checking, real-world, and synthetic graphs shown in Table~\ref{Table:graph}. 
\begin{itemize}
    \item 
The \textit{cambridge.6}, \textit{bakery.6} and \textit{leader-filters.7} graphs come from the model checking problems in the BEEM database~\cite{pelanek2007beem}, which are implicit and can be generated on-the-fly. For convenience, these graphs are converted to explicit graphs~\cite{bloemen2016multi} and stored in files. 
\item
The \textit{livej}, \textit{patent}, and \textit{wikitalk} graphs are obtained from SNAP~\cite{snapnets} \footnote{\url{https://snap.stanford.edu}}; they represent the Live-Journal social network~\cite{backstrom2006group}, the U.S patent dataset is maintained by the National Bureau of Economic Research~\cite{leskovec2005graphs}, and Wikipedia Talk (communication) network~\cite{leskovec2010signed}, respectively.
The \textit{dbpedia}, \textit{baidu}, and \textit{wiki-talk-en} graphs are collected from the University of Koblenz-Landau~\cite{kunegis2013konect}; they represent the DBpedia network~\cite{konect:dbpedia2}, the hyperlink network between the articles of the Baidu~\cite{konect:zhishi} encyclopedia, and the communication network of the English Wikipedia~\cite{konect:sun_2016_49561}, respectively.
\item
The \textit{com-friendster}, \textit{twitter} and \textit{twitter-mpi} are three super large graphs with billions of edges obtained from the Network Repository \cite{nr} \footnote{\url{http://networkrepository.com}}; they represent the online gaming social network \cite{friendster}, the follower network from Twitter \cite{cha2010measuring}, the twitter follow data collected in 2009 \cite{icwsm10cha}, respectively.
\item
The \textit{ER}, \textit{BA}, and \textit{RMAT} graphs are synthetic graphs; they are generated by the SNAP~\cite{snapnets} system using the Erd\"{o}s-R\'enyi graph model (which generates a random graph), the Barabasi-Albert graph model (which generates a graph with power-law degree distribution), and the R-MAT graph model (which generates large-scale realistic graph  similar to social networks), respectively; for these generated graphs, the average degree is fixed to $8$ by choosing $1,000,000$ vertices and $8,000,000$ edges. 
\end{itemize}
All these graphs are stored in the Compressed Sparse Row (CSR) binary format~\cite{Hong2012,hong2013fast}, which is compact and memory bandwidth-friendly. Taking the super large graph \textit{twitter} for example, the text file that includes all edges requires 30 GB while the CSR binary format only requires 6 GB.

\begin{table*}[!htb]
\footnotesize
\centering

\begin{tabular}{l|r r r r r r}
\toprule
Name & $|V|$ & $|E|$ & $\it{Deg_{in}}$ & $\it Deg_{out}$& $\alpha$ &  \%Trim\\
  \midrule
cambridge.6         & 3.3M	    &9.5M		&15	    &6          &65	        &  0.25\%   \\
bakery.6            & 11.8M	    &40.4M		&24	    &4          &47	        &  22.26\%  \\
leader-filters.7    & 26.3M	    &91.7M	    &12	    &6          &73	        &  100.00\%\\ \hline
dbpedia             & 4.0M	    &13.8M	    &473.0K   &1.0K       &116        &  36.23\%  \\
baidu               & 2.1M	    &17.8M	    &98.0K	&2.6        &9	        &  27.97\%  \\
livej               & 4.8M	    &69.0M	    &14.0K	& 20.3K     &8	        &  12.23\%  \\
patent              & 6.0M	    &16.5M	    &779    &770        &5	        &  100.00\%  \\
wiki-talk-en        & 3.0M	    &25.0M	    &121.3K   & 488.2K    &7	        &  87.42\%  \\
wikitalk            & 2.4M      &5.0M       &3.3K   & 100.0K    &5          &   94.49\%\\ \hline
com-friendster      & 125M      &1.8B       &4.2K   &3.6K       & 11.7K    & 100.00\% \\
twitter             &41.4B      &1.4B       &770.2K &3.9M       &6          & 10.05\% \\
twitter-mpi         &52.6B      &2.0B       &3.5M   &780.0K     &7          &17.52\%\\ \hline
ER                  & 1.0M	    &8.0M	    &25	    & 24        &3	        &   0.03\% \\
BA                  & 1.0M	    &8.0M	    &8	    &5.2K       &122	    & 100\%\\
RMAT                & 1.0M	    &8.0M	    &335	& 1.9K      &7.0K       & 99.98\%\\
  \bottomrule
\end{tabular} 

\caption{The characteristics for model checking, real-world, and synthetic graphs. Here, columns denote the number of vertices $n$, the number of edges $m$, the maximum in-degree, the maximum out-degree, the number of peeling steps, and the percentage of trimmable vertices, respectively.}
\label{Table:graph}
\end{table*}

Table~\ref{Table:graph} provides an overview of the 15 tested graphs. For some graphs, e.g.~\textit{cambridege.6} and \textit{ER}, less than 1\% of vertices can be trimmed. However, for most of the other graphs, a high ratio of vertices can be trimmed, especially for \textit{leader-filters.7}, \textit{BA}, and \textit{com-friendster}, whose vertices can be trimmed altogether.  
More importantly, for most graphs, the trimming steps $\alpha$, maximum in-degree $\Degin$, and maximum out-degree $\Degout$ are always small. When analyzing the parallel time complexity, these three values are associated with the parallel depths. The small values of the depths indicate that the execution can be highly parallelized \cite{blelloch2010parallel}.

\subsection{Workload Balance}
Given a tested graph, all vertices are partitioned into multiple chunks, which can be dynamically assigned to workers for workload balance. 
One issue is how to determine the size of chunks. A large size of chunks may lead to workload imbalance, while a small size of chunks may lead to a high cost of scheduling. 
Since the trend is similar for all tested graphs, we select three typical graphs with a variety of $\Degin$, $\Degout$, and $\alpha$ for the evaluation.
In Figure \ref{Figure:trunk-size}, we test three trimming algorithms over three selected graphs, \textit{leader-filters.7}, \textit{livej}, and \textit{wiki-talk-en}, that have millions of vertices, by using $16$ workers and varying the chunk size from $1$ to $2^{20}$. All three trimming algorithms tend to be efficient when choosing a chunk size between $2^{10}$ and $2^{16}$. 
Therefore, in our experiments, we fix the chunk size to $2^{12}=4096$ for both workload balance and efficient scheduling. 


\iftrue
\def\fwide{0.3}
\begin{figure*}[!htb]
    \centering
    \begin{subfigure}[b]{1\textwidth}
    \centering
    \footnotesize
    \begin{tikzpicture}
        \begin{customlegend}[legend columns=5,legend style={align=center,draw=none,column sep=2ex},
                legend entries={{AC3Trim} ,
                                {AC4Trim} ,
                                {AC6Trim} 
                                }]
            \addlegendimage{color=red, mark=triangle*,solid}
            \addlegendimage{color=darkgreen, mark=square*,solid}
            \addlegendimage{color=blue, mark=oplus*, solid}
        \end{customlegend}
    \end{tikzpicture}
    \end{subfigure}
    \def\yminmax{}

       \def\csvfilename{leader_filters.7--.csv}
    \def\title{leader-filters.7}
     \TRUNKSUBFIGURE
    \def\csvfilename{livej--.csv}
    \def\title{livej}
     \TRUNKSUBFIGURE
    \def\csvfilename{wiki_talk_en--.csv}
    \def\title{wiki-talk-en}
     \TRUNKSUBFIGURE

    \caption{The practical running time for \emph{AC3Trim, AC4Trim} and \emph{AC6Trim} with 16 workers by varying the chunk size.}
    \label{Figure:trunk-size}
\end{figure*}
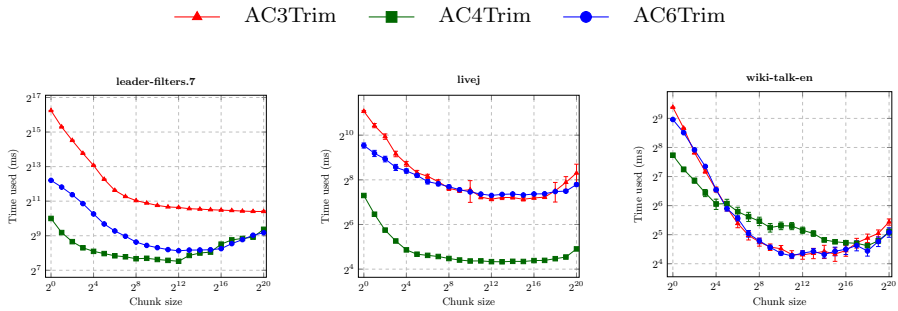
\fi

The other issue is the upper-bound size of the waiting set $Q_p$ for each worker $p$ in \emph\textit{AC4Trim} and \emph{AC6Trim}. 
Here, $Q_p$ is private to worker $p$, and thus the vertices in $Q_p$ are processed sequentially by worker $p$ without synchronization. A large size of $Q_p$ may lead to workload imbalance. 
In Table \ref{table:maxpush}, over all tested graphs we show the upper-bound size of $Q_p$, denoted as $|Q_p|$, for \emph{AC4Trim} and \emph{AC6Trim} by using $16$ workers. 
We can see that $|Q_p|$ is relatively small compared with millions of vertices. 
Further, over all tested graphs \emph{AC6Trim} has $|Q_p|$ bounded by 900, while \emph{AC4Trim} has $|Q_p|$ up to 20761. Especially, \emph{AC6Trim} has much smaller values of $|Q_p|$ than \emph{AC4Trim} for graphs like \textit{dbpedia} and \textit{twitter-mpi}. That means that \emph{AC6Trim} on average has better workload balance than \emph{AC4Trim}.

\begin{table*}[!htb]
\footnotesize
\centering
 \begin{tabular}{l | r  r}%
 \toprule
    Name & \emph{AC4Trim} $|Q_p|$ & \emph{AC6Trim} $|Q_p|$  \\ 
    \midrule
cambridge.6 & 17 & 6 \\
bakery.6 & 21 & 52 \\
leader-filters.7 & 16 & 103 \\
dbpedia & \textbf{20761} & \textbf{852} \\
baidu & 439 & 108 \\
livej & 274 & 8 \\
patent & 95 & 55 \\
wiki-talk-en & 84 & 5 \\
wikitalk & 33 & 7 \\
com-friendster & 646 & 677 \\
twitter & 293 & 287 \\
twitter-mpi & 15217 & 327 \\
ER & \textbf{1} & \textbf{1} \\
BA & 66 & 27 \\
RMAT & 299 & 411 \\
    \bottomrule
\end{tabular}
\caption{The upper-bound size of $Q_p$ for \textit{AC4Trim} and \textit{AC6Trim} by using $16$ workers. The best and worst cases are in \textbf{bold} for each column.}
\label{table:maxpush}
\end{table*}

\subsection{Evaluating the Number of Traversed Edges}
To evaluate the Arc-Consistency algorithms, the traditional approach is to count the total number of checked constraints. For each constraint check, a pair of values in the domain $D(X_i)$ and $D(X_j)$ is checked. Such an evaluation is reasonable because 1) most of the running time is spent on checking numerous constraints, 2)~the time used for checking each constraint significantly varies for different kinds of arc-consistency problems.
Analogous to evaluating Arc-Consistency algorithms, we compare the total number of traversed edges of three trimming algorithms. This is especially meaningful for the implicit graphs since their edges are generated on-the-fly, costing most of the running time. 

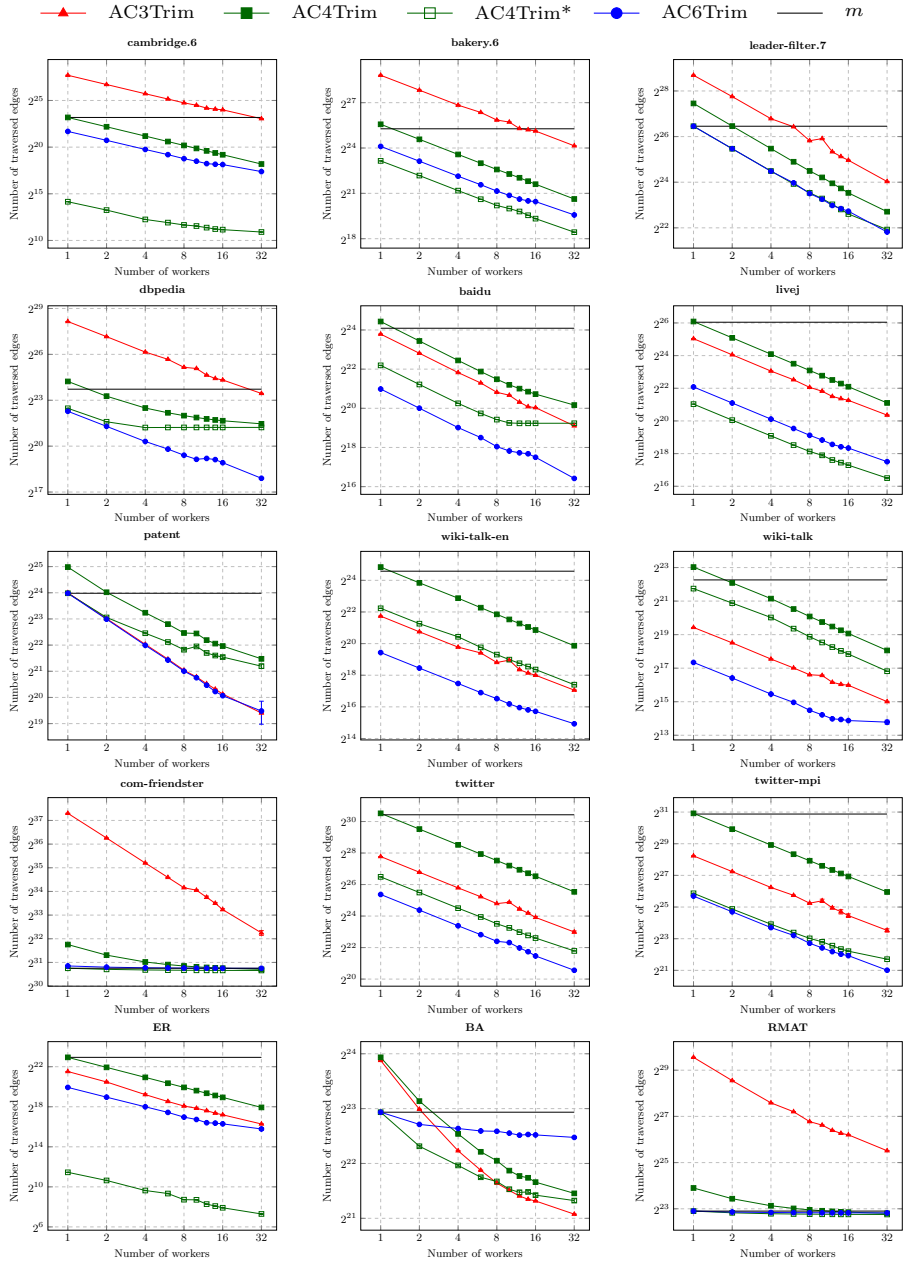
\begin{figure*}[!htb]
    \centering
    \begin{subfigure}[b]{1\textwidth}
    \centering
    \scriptsize
    \begin{tikzpicture}
        \begin{customlegend}[legend columns=5,legend style={align=center,draw=none,column sep=2ex},
                legend entries={{AC3Trim} ,
                                {AC4Trim} ,
                                {AC4Trim*},
                                {AC6Trim} ,
                                {$m$}
                                }]
            \addlegendimage{color=red, mark=triangle*,solid}
            \addlegendimage{color=darkgreen, mark=square*,solid}
            \addlegendimage{color=darkgreen, mark=square,solid}
            \addlegendimage{color=blue, mark=oplus*, solid}
            \addlegendimage{color=black, solid}
        \end{customlegend}
    \end{tikzpicture}
    \end{subfigure}
     \def\yminmax{}
    \def\csvfilename{cambridge.6-.csv}
    \def\title{cambridge.6}
    \NUMSUBFIGURE
    \def\csvfilename{bakery.6-.csv}
    \def\title{bakery.6}
     \NUMSUBFIGURE
    \def\csvfilename{leader_filters.7-.csv}
    \def\title{leader-filter.7}
     \NUMSUBFIGURE
    \def\csvfilename{dbpedia-.csv}
   \def\title{dbpedia}
     \NUMSUBFIGURE
     \def\csvfilename{baidu-.csv}
    \def\title{baidu}
     \NUMSUBFIGURE
         \def\csvfilename{livej-.csv}
   \def\title{livej}
     \NUMSUBFIGURE
         \def\csvfilename{patent-.csv}
    \def\title{patent}
     \NUMSUBFIGURE
         \def\csvfilename{wiki_talk_en-.csv}
    \def\title{wiki-talk-en}
     \NUMSUBFIGURE
         \def\csvfilename{wikitalk-.csv}
   \def\title{wiki-talk}
     \NUMSUBFIGURE

         \def\csvfilename{com-friendster-.csv}
   \def\title{com-friendster}  
     \NUMSUBFIGURE  
         \def\csvfilename{twitter-.csv}
   \def\title{twitter}
     \NUMSUBFIGURE
               \def\csvfilename{twitter_mpi-.csv}
   \def\title{twitter-mpi}
     \NUMSUBFIGURE

         \def\csvfilename{er-1m-8m-.csv}
     \def\title{ER}
     \NUMSUBFIGURE
         \def\csvfilename{ba-1m-8m-.csv}
    \def\title{BA}
     \NUMSUBFIGURE
         \def\csvfilename{rmat-1m-8m-.csv}
   \def\title{RMAT}
     \NUMSUBFIGURE
    \caption{The maximum number of traversed edges per worker for \emph{AC3Trim}, \emph{AC4Trim}, \emph{AC4Trim*} and \emph{AC6Trim} by varying the number of workers. The number of edges $m$ in a graph is chosen as a baseline.}
    \label{figure:edge}
\end{figure*}

In this experiment, we exponentially increase the number of workers from 1 to 32 and count the largest number of traversed edges per worker over graphs in Table \ref{Table:graph}. The plots in Figure \ref{figure:edge} depict the maximum number of traversed edges per worker for the three compared methods. The x-axis is the number of workers and the y-axis is the number of traversed edges.
Also, we choose the total number of edges in a graph as a baseline (denoted as $m$).
Note that, for the AC-4-based trimming algorithm, the out-degree counter of each vertex $v$ in graph is initialized as $v.\degout = |v.\post|$. 
To calculate $v.\degout$,  there are two cases: 1) we can traverse all $v$'s successors one by one to count the total numbers of successors (denoted as \emph{AC4Trim}), which means all $v$'s edges of are traversed once;
2) if $v'$ successors are stored successively, we can take the difference between the index of $v$'s first successor and $v$'s last successor without traversing the edges (denoted as \emph{AC4Trim*}), which means only $v$ is traversed once and all $v$'s edges are not traversed. Absolutely, \emph{AC4Trim} traverses a higher number of edges than \emph{AC4Trim*}.

\def\fwide{0.3}

In Figure \ref{figure:edge}, a first look over nearly all testing graphs reveals that the number of traversed edges of all three algorithms is linearly decreasing with an increasing number of workers, which achieves a good load balance.
Over most of the testing graphs, \emph{AC6Trim} traverses fewer edges compared with \emph{AC4Trim} and \emph{AC3Trim}. \emph{AC3Trim} sometimes traverses more edges than the baseline $m$ for some graphs with large $\alpha$. Specifically, we make four observations:

- Over graphs with a higher value of $\alpha$, e.g. \emph{cambridge.6, bakery.6, leader-filter.7, dbpedia, com-friendster} and \emph{RMAT}, \emph{AC3Trim} always traverses much more edges than both \emph{AC4Trim} and \emph{AC6Trim} and even more than the baseline $m$. This is because \emph{AC3Trim} has the worst work complexity $\mathcal{O}(\alpha(n+m))$ which requires $\alpha$ number of repetitions, but \emph{AC4Trim} and \emph{AC6Trim} have a linear work complexity of $\mathcal{O}(n+m)$.

- Over the graphs with a lower value of $\alpha$, e.g. \emph{wiki-talk-en} and \emph{wikitalk}, \emph{AC3Trim} traverses fewer edges than \emph{AC4Trim}. This is because \emph{AC3Trim} executes always close to the best-case time complexity, but \emph{AC4} executes always close to the worst-case time complexity. This is why \emph{AC3Trim} is sometimes more powerful than \emph{AC4Trim} in real-world graphs with a relatively low value of $\alpha$.

- Over all graphs, \emph{AC6Trim} always traverses much fewer edges than \emph{AC4Trim} even if they have nearly the same time complexity. The reason is that \emph{AC6Trim} can traverse only part of the edges of removed vertices, which is close to the best-case time complexity. However, \emph{AC4Trim} has to traverse all edges to initialize the counters $\forall v\in V: v.\degout$ and all ingoing edges of removed vertices, which is close to the worst-case time complexities. Therefore, \emph{AC6Trim} certainly traverses fewer edges than \emph{AC4Trim}.

- Over all graphs, for all three methods, the number of traversed edges is well bounded without obvious variation even these three methods are non-deterministic. That is, for the number of traversed edges, the affect of non-determinism can be omitted.

\begin{table}[!htb]
\small
\centering
\begin{tabular}{l|r r r | r r }
 \toprule
        & \multicolumn{3}{c|}{1-worker vs 16-worker} & AC3Trim vs& AC4Trim vs  \\
 Name   & AC3Trim & AC4Trim & AC6Trim & AC6Trim & AC6Trim \\
  \midrule
  
cambridge.6     & 13.04 & 15.97 & 11.74 & \textbf{58.29} & 2.08 \\
bakery.6        & 12.90 & 15.61 & 12.58 & 25.44 & 2.22 \\
leader-filters.7        & 13.22 & 15.15 & 13.23 & 4.71  & 1.75 \\
dbpedia & 14.13 & 5.94  & 10.26 & 42.43 & 6.69 \\
baidu   & 13.49 & 13.07 & 11.19 & 5.79  & 9.35 \\
livej   & 13.69 & 15.90 & 13.38 & 7.58  & 13.51 \\
patent  & 14.63 & 8.09  & 15.07 & 1.04  & 3.72 \\
wiki-talk-en    & 13.33 & 15.76 & 13.17 & 4.87  & 35.33 \\
wikitalk        & 10.89 & 15.55 & 11.05 & 4.31  & \textbf{36.51} \\
com-friendster  & 16.78 & \textbf{2.00}  & 1.07  & 5.57  & 1.00 \\
twitter & 14.54 & 15.92 & \textbf{14.95} & 5.45  & 33.31 \\
twitter-mpi     & 13.66 & 15.87 & 13.56 & 5.77  & 32.00 \\
ER      & \textbf{20.00} & \textbf{16.00} & 12.47 & 1.87  & 6.24 \\
BA      & \textbf{5.90}  & 4.84  & 1.33  & \textbf{0.43}  & \textbf{0.55} \\
RMAT    & 10.23 & 2.07  & \textbf{1.05}  & 10.39 & 1.02 \\

\bottomrule
\end{tabular}
\caption{Compare the ratio for the maximum number of traversed edges per worker. The best and worst cases are in \textbf{bold} for each column.}
\label{table:edgespeedup}
\end{table}

In Table \ref{table:edgespeedup}, columns 2 - 4 compare the ratio of the maximum number of traversed edges per worker between using a single worker and using 16 workers for \emph{AC3Trim}, \emph{AC4Trim} and \emph{AC6Trim}, respectively. We can see that for \emph{AC3Trim}, the ratio is at least 5.9 as \emph{AC3Trim} is easy to be parallelized without using locks or atomic primitives. We also can see that for \emph{AC3Trim} the ratio is larger than 16 in some graphs, e.g. \emph{com-friendster} and \emph{ER}. The reason is that parallel \emph{AC3Trim} is non-deterministic; that is, different trimming orders lead to different numbers of traversed edges; if numerous vertices are early determined as \DEAD, the time complexity is close to the best case. 
For \emph{AC4Trim} and \emph{AC6Trim}, the ratio is relatively low in some graphs with large $\alpha$, e.g. \emph{RMAT} and \emph{com-friendster}, as large $\alpha$ always leads to high working depths.

In columns 5 and 6 of Table \ref{table:edgespeedup}, we fix using 16 workers and compare the ratio of traversed edge numbers between \emph{AC3Trim} and \emph{AC6Trim} and between \emph{AC4Trim} and \emph{AC6Trim}. We can see that \emph{AC6Trim} traverses much fewer edges than \emph{AC3Trim}, up to 58 times over the graph \emph{cambridge.6}; \emph{AC6Trim} traverses much fewer edges then \emph{AC4Trim}, up to 36 times over the graph \emph{wikitalk}. 
\emph{AC3Trim} traverses the fewest edges in some graphs, e.g. \emph{BA}, as the time complexity is close to the best case.

\subsection{Evaluating the Real Running Time}
In this experiment,  we exponentially increase the number of workers from 1 to 32 and evaluate the real running time over graphs in Table~\ref{Table:graph}. The plots in Figure~\ref{Figure:time} depict the performance of the three compared methods. The x-axis is the number of workers and the y-axis is the execution time (millisecond). 
The first look over all testing graphs reveals that the trends for the running time are much different from the trends for the number of traversed edges shown in Figure~\ref{figure:edge}. This can be explained as follow:

 - Although \emph{AC6Trim} always traverses the fewest numbers of edges, \emph{AC6Trim} is slower than \emph{AC4Trim} in some graphs, e.g. \emph{cambridge.6, livej, pokec} and \emph{ER}, and even \emph{AC3Trim} is the fastest in some graphs, e.g. \emph{BA}. The main reason is that \emph{AC6Trim} is cache-unfriendly while \emph{AC3Trim} and \emph{AC4Trim} are cache-friendly. That means \emph{AC6Trim} cannot fully use caches to archive the best performance even if \emph{AC6Trim} traverse the least number of edges. The other reason is that maintaining the supporting sets in \emph{AC6Trim} costs much more computational time than maintaining the out-degree counters in \emph{AC4Trim}; there is no auxiliary data structure in \emph{AC3Trim} so that no computational time is spent on this part. 
 
- In \emph{AC4Trim}, the running times have a wide variation in certain graphs, e.g. \emph{bakery.6}, \emph{leader-filter.7}, \emph{livej}. The reason is that \emph{AC4Trim} is sometimes less cache-friendly. The unexpected missing cache leads to the performance decreased. However, \emph{AC3Trim} is always cache-friendly and \emph{AC6Trim} is always cache-unfriendly so that their performance is more stable than \emph{AC3Trim}. 

- In \emph{AC6Trim}, the running times begin to increase when using more than 4 workers in certain graphs, e.g. \emph{dbpedia} and \emph{baidu}. The reason is that the supporting set shared by multi-worker is synchronized by busy waiting, which leads to contention. At the same time, \emph{AC4Trim} still has not obvious speedup as there is less contention to use atomic primitive updating the out-degree counters. 
However, \emph{AC3Trim} always has a speedup by multiple workers and even has the best performance with 16 workers in some graphs, e.g.~ \emph{BA}, as \emph{AC3Trim} has no shared data structures and thus no contention.  

\iftrue
\def\fwide{0.3}
\begin{figure*}[!htb]
    \centering
    \begin{subfigure}[b]{1\textwidth}
    \centering
    \footnotesize
    \begin{tikzpicture}
        \begin{customlegend}[legend columns=5,legend style={align=center,draw=none,column sep=2ex},
                legend entries={{AC3Trim} ,
                                {AC4Trim} ,
                                {AC6Trim} 
                                }]
            \addlegendimage{color=red, mark=triangle*,solid}
            \addlegendimage{color=darkgreen, mark=square*,solid}
            \addlegendimage{color=blue, mark=oplus*, solid}
           
        \end{customlegend}
    \end{tikzpicture}
    \end{subfigure}
    \def\yminmax{}
    \def\csvfilename{cambridge.6.csv}
    \def\title{cambridge.6}
    \TIMESUBFIGURE
    \def\csvfilename{bakery.6.csv}
    \def\title{bakery.6}
     \TIMESUBFIGURE
    \def\csvfilename{leader_filters.7.csv}
    \def\title{leader-filter.7}
     \TIMESUBFIGURE
    \def\csvfilename{dbpedia.csv}
   \def\title{dbpedia}
     \TIMESUBFIGURE
     \def\csvfilename{baidu.csv}
    \def\title{baidu}
     \TIMESUBFIGURE
         \def\csvfilename{livej.csv}
   \def\title{livej}
     \TIMESUBFIGURE
         \def\csvfilename{patent.csv}
    \def\title{patent}
     \TIMESUBFIGURE
         \def\csvfilename{wiki_talk_en.csv}
    \def\title{wiki-talk-en}
     \TIMESUBFIGURE
         \def\csvfilename{wikitalk.csv}
   \def\title{wiki-talk}
     \TIMESUBFIGURE

         \def\csvfilename{com-friendster.csv}
   \def\title{com-friendster} \def\yminmax{ymin=20000, ymax=80000,}
     \TIMESUBFIGURE \def\yminmax{}
         \def\csvfilename{twitter.csv}
   \def\title{twitter}
     \TIMESUBFIGURE
               \def\csvfilename{twitter_mpi.csv}
   \def\title{twitter-mpi}
     \TIMESUBFIGURE

         \def\csvfilename{er-1m-8m.csv}
     \def\title{ER}
     \TIMESUBFIGURE
         \def\csvfilename{ba-1m-8m.csv}
    \def\title{BA}
     \TIMESUBFIGURE
         \def\csvfilename{rmat-1m-8m.csv}
   \def\title{RMAT}
     \TIMESUBFIGURE

    \caption{The real running time for \emph{AC3Trim, AC4Trim} and \emph{AC6Trim} by varying the number workers.}
    \label{Figure:time}
\end{figure*}
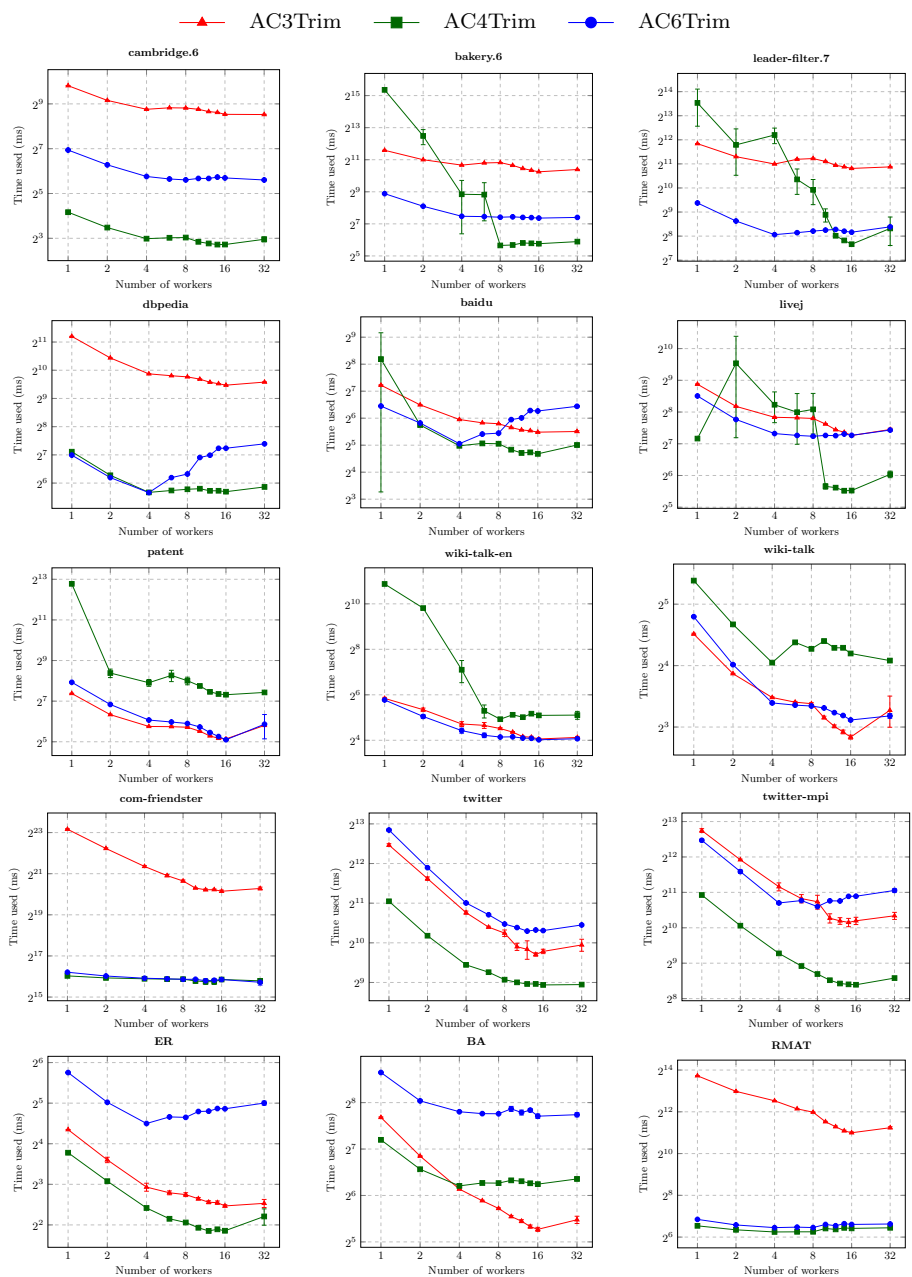

\fi 


\begin{table}[!htb]
\small
\centering
\begin{tabular}{l|r r r | r r }
 \toprule
        & \multicolumn{3}{c}{16-workers speedup vs 1-worker} & \multicolumn{2}{|c}{AC6Trim speedup vs}  \\
 Name   & AC3Trim & AC4Trim & AC6Trim & AC3Trim & AC4Trim \\
  \midrule
cambridge.6     & 2.42  & 2.72  & 2.37  & 7.15  & 0.13 \\
bakery.6        & 2.54  & \textbf{771.96}        & 2.87  & 7.39  & 0.33 \\
leader-filters.7        &\textbf{ 2.04}  & 58.42 & 2.32  & 6.27  & 0.71 \\
dbpedia & 3.31  & 2.67  & \textbf{0.84}  & 4.71  & 0.34 \\
baidu   & 3.34  & 11.38 & 1.14  & 0.58  & 0.33 \\
livej   & 3.07  & 3.11  & 2.36  & 1.00  & 0.30 \\
patent  & 4.71  & 44.02 & \textbf{7.11}  & 1.03  & \textbf{4.65} \\
wiki-talk-en    & 3.45  & 54.85 & 3.37  & 1.02  & 2.10 \\
wikitalk        & 3.20  & 2.28  & 3.21  & 0.82  & 2.12 \\
com-friendster  & 8.09  & 1.13  & 1.28  & 19.62 & 1.00 \\
twitter & 6.42  & 4.32  & 5.80  & 0.70  & 0.39 \\
twitter-mpi     & 5.85  & 5.79  & 2.99  & 0.62  & 0.18 \\
ER      & 3.68  & 3.79  & 1.86  & 0.19  & \textbf{0.12} \\
BA      & 5.32  & 1.94  & 1.92  & \textbf{0.18}  & 0.36 \\
RMAT    & \textbf{6.62}  & \textbf{1.09}  & 1.18  & \textbf{20.97} & 0.88 \\

 \bottomrule
\end{tabular}
\caption{Compare the speedups for running times between using 1-worker and 16-worker for \emph{AC3Trim}, \emph{AC4Trim}, and \emph{AC6Trim}, respectively; by fixing with 16 workers, compare the running time speedup between \emph{AC6Trim} and \emph{AC3Trim} and between \emph{AC6Trim} and \emph{AC4Trim}. The best and worst cases are in \textbf{bold} for each column.}
\label{table:timepeedup}
\end{table}

In columns 2 - 4 of Table \ref{table:timepeedup} we compare the running time speedup between using one worker and 16 workers for \emph{AC3Trim}, \emph{AC4Trim} and \emph{AC6Trim}, respectively. It is clear that \emph{AC3Trim} achieves the best speedup and \emph{AC4} achieves the worst speedup. This is because of the contention on shared data structures with multiple workers. In columns 5 and 6 of Table \ref{table:timepeedup} we fix using 16 workers and compare the speedups for the running time between \emph{AC6Trim} and \emph{AC3Trim} and between \emph{AC6Trim} and \emph{AC4Trim}. We can see that our \emph{AC6Trim} is up to 24 times faster than \emph{AC3Trim} over \emph{RMAT} and up to 7.8 times faster than \emph{AC4Trim} over \emph{leader-filters.7}. However, in some graphs, \emph{AC4Trim} and \emph{AC3Trim} have better performances than \emph{AC6Trim}.

\subsection{Evaluating Stability }
One issue is the stability of the trimming algorithms when executing the same algorithm multiple times. 
In this experiment, we compare 50 testing result over three chosen graphs, \textit{leader-filters.7},\textit{livej}, and \textit{wiki-talk-en}. In Figure~\ref{Figure:stable}, the x-axis of plots is the index of the repeating times.
The upper three plots in Figure~\ref{Figure:stable} depict the number of traversed edges for three trimming methods, in which the y-axis is the number of traversed edges. We observe that the number of traversed edges is well bounded for all three trimming methods. 
The lower three plots in Figure \ref{Figure:stable} depict the running time for three trimming methods, in which the y-axis is the running time. We observe that \emph{AC4Trim} always has a wider variation than other methods. 
The reason is that parallel \emph{AC4Trim} is non-deterministic, which means each time the order of removed vertices is different; \emph{AC4Trim} is not always cache-friendly as vertices are not sequentially traversed; there is a high probability that the performance decreases due to the unexpected missing cache. 
Even \emph{AC3Trim} and \emph{AC6Trim} are also non-deterministic, \emph{AC3Trim} is cache-friendly and \emph{AC6Trim} is cache-unfriendly; cache-friendliness does not always lead to a wide performance variation.



\iftrue
\def\fwide{0.3}
\begin{figure*}[!htb]
    \centering
    \begin{subfigure}[b]{1\textwidth}
    \centering
    \footnotesize
    \begin{tikzpicture}
        \begin{customlegend}[legend columns=5,legend style={align=center,draw=none,column sep=2ex},
                legend entries={{AC3Trim} ,
                                {AC4Trim} ,
                                {AC6Trim} 
                                }]
            \addlegendimage{color=red, mark=triangle*,solid}
            \addlegendimage{color=darkgreen, mark=square*,solid}
            \addlegendimage{color=blue, mark=oplus*, solid}
           
        \end{customlegend}
    \end{tikzpicture}
    \end{subfigure}
    \def\yminmax{}    
    \def\myylabel{Number of traversed edges}
    
    \def\csvfilename{leader_filters.7-snum.csv}
    \def\title{leader-filters.7}
    \STABILITYFIGURE
    \def\csvfilename{livej-snum.csv}
    \def\title{livej}
     \STABILITYFIGURE
    \def\csvfilename{wiki_talk_en-snum.csv}
    \def\title{wiki-talk-en}
     \STABILITYFIGURE
    \def\myylabel{Time used (ms)}
    
    \def\csvfilename{leader_filters.7-stime.csv}
    \def\title{leader-filters.7}
    \STABILITYFIGURE
    \def\csvfilename{livej-stime.csv}
    \def\title{livej}
     \STABILITYFIGURE
    \def\csvfilename{wiki_talk_en-stime.csv}
    \def\title{wiki-talk-en}
     \STABILITYFIGURE
    \caption{The stability of the traversed edge number and the running time for \emph{AC3Trim, AC4Trim} and \emph{AC6Trim}.}
    \label{Figure:stable}
\end{figure*}
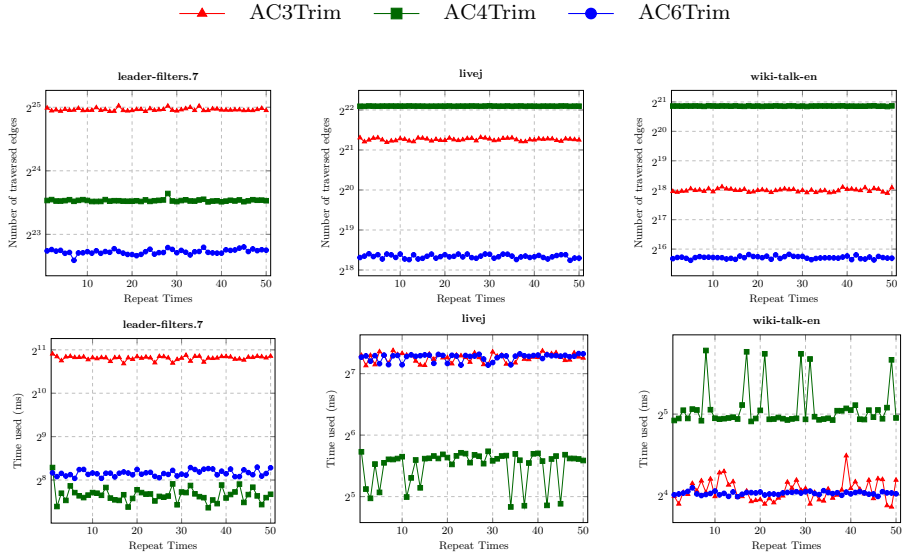
\fi

\subsection{Evaluating Scalability}
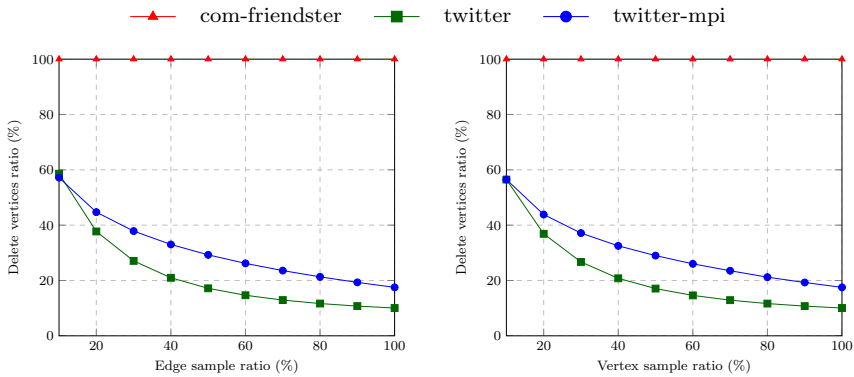
\begin{figure*}[!htb]
    \centering
     \footnotesize
     \begin{subfigure}[b]{1\textwidth}
    \centering
    \footnotesize
    \begin{tikzpicture}
        \begin{customlegend}[legend columns=5,legend style={align=center,draw=none,column sep=2ex},
                legend entries={{com-friendster} ,
                                {twitter} ,
                                {twitter-mpi} 
                                }]
            \addlegendimage{color=red, mark=triangle*,solid}
            \addlegendimage{color=darkgreen, mark=square*,solid}
            \addlegendimage{color=blue, mark=oplus*, solid}
           
        \end{customlegend}
    \end{tikzpicture}
    \end{subfigure}
    
    \def\myxlabel{Edge sample ratio (\%)}
    \def\csvfilename{deletenumedge.csv}
    \DELETENUMBAR
    \def\myxlabel{Vertex sample ratio (\%)}
    \def\csvfilename{deletenumver.csv}
     \DELETENUMBAR
  
    \caption{The ratio of trimmable vertices.}
        
    \label{fig:trim-ratio}
\end{figure*}

An issue is the scalability of the trimming algorithms when the size of graphs is varied. 
In this experiment, we test the scalability of the trimming algorithms over the three largest graphs, i.e., \textit{com-friendster, twitter}, and \textit{twitter-mpi}. 
Using 16 workers, we vary the number of edges and vertices by randomly sampling at a ratio from 10\% to 100\%, respectively. By sampling the edges, we simply remove the unsampled edges. By sampling the vertices, we simply set the unsampled vertices to \texttt{DEAD}.
As shown in Figure \ref{fig:trim-ratio}, we can see that the smaller ratio of sampling edges or vertices always leads to the higher ratio of trimable vertices, e.g. \textit{twitter} and \textit{twitter-mpi}; but for \textit{com-friendster} all vertices are always trimmable with any ratio of sampling. Especially when sampling 10\% edges or vertices, nearly 60\% of vertices can be trimmed for \textit{twitter} and \textit{twitter-mpi}, and without sampling less than 20\% of vertices can be trimmed. This is result is reasonable as more unsampled edges or vertices will lead to more vertices without out-going edges and thus can be trimmed.  


\def\fwide{0.45}

We show the result of sampling edges in Figure \ref{fig:sample-edge}, in which the x-axis of plots is the ratio of sampled edges. The upper three plots in Figure \ref{fig:sample-edge} depict the maximum number of traversed edges per worker. 
We observe that the number of traversed edges is generally increasing with the ratio of the sampled edges. Not surprisingly, \emph{AC6Trim} traverses the least number of edges, and \emph{AC3Trim} traverses the highest number of edges. 
But for \emph{AC3Trim} the number of traversed edges fluctuates when increasing the sampling ratio of edges as the number of peeling steps $\alpha$ may fluctuate with a different sampling ratios of edges.  
The lower three plots in Figure \ref{fig:sample-edge} depict the real running time. We make three observations.

- Over \textit{com-friendster}, \emph{AC6Trim} has the best performance. The reason is that 100\% of vertices can be trimmed so that \emph{AC4Trim} accesses all vertices almost randomly. In this case, \emph{AC4Trim} is likely too cache-unfriendly, and the cache can not provide an obvious speedup. 

- Over \textit{twitter} and \textit{twitter-mpi}, \emph{AC4Trim} has a wide variation and \emph{AC4Trim} performs worse than \emph{AC6Trim} in most of cases. The reason is that for \emph{AC4Trim} more trimmable vertices lead to the cache being less effective, and sometimes the cache can provide a speedup but sometimes not; but \emph{AC6Trim} is cache-unfriendly, and the cache cannot affect the running time. 

- Over \textit{twitter} and \textit{twitter-mpi}, \emph{AC3Trim} always has as good performance as \emph{AC6Trim} even if \emph{AC3Trim} traverse much more edges than \emph{AC6Trim}. The reason is that \emph{AC3Trim} is cache-friendly and achieve a high speedup with caching.

\iftrue
    \def\fwide{0.3}
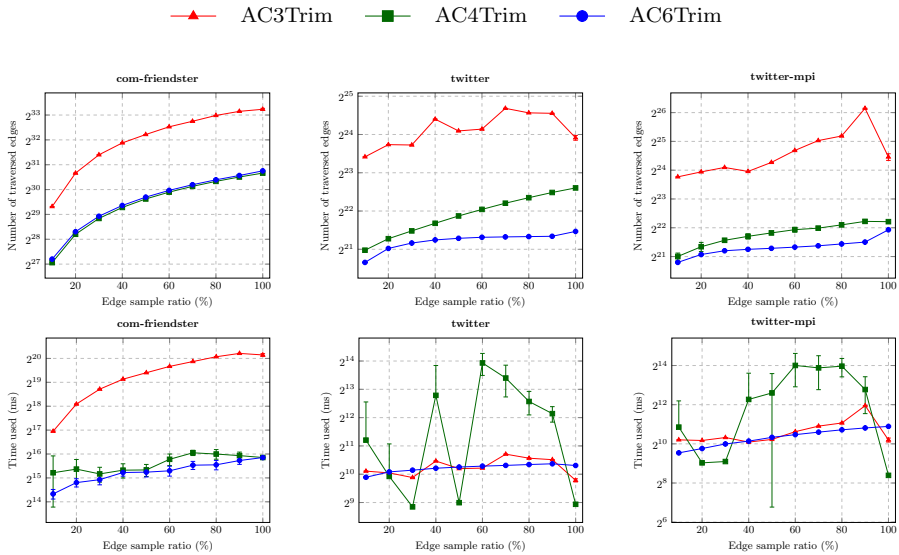
\begin{figure*}[!htb]
    \centering
    \begin{subfigure}[b]{1\textwidth}
    \centering
    \footnotesize
    \begin{tikzpicture}
        \begin{customlegend}[legend columns=5,legend style={align=center,draw=none,column sep=2ex},
                legend entries={{AC3Trim} ,
                                {AC4Trim} ,
                                {AC6Trim} ,
                                }]
            
            \addlegendimage{color=red, mark=triangle*,solid}
            \addlegendimage{color=darkgreen, mark=square*,solid}
            \addlegendimage{color=blue, mark=oplus*, solid}
            
        \end{customlegend}
    \end{tikzpicture}
    \end{subfigure}
    \def\yminmax{}
    
     \def\myxlabel{Edge sample ratio (\%)}
    \def\myylabel{Number of traversed edges}
   
    \def\csvfilename{com-friendster-esnum.csv}
    \def\title{com-friendster}
    \SAMPLEFIGURE 
    \def\csvfilename{twitter-esnum.csv}
    \def\title{twitter}
    \SAMPLEFIGURE 
    \def\csvfilename{twitter_mpi-esnum.csv}
    \def\title{twitter-mpi}
    \SAMPLEFIGURE 

     \def\myylabel{Time used (ms)}
    \def\csvfilename{com-friendster-estime.csv}
    \def\title{com-friendster}
    \SAMPLEFIGURE
    \def\csvfilename{twitter-estime.csv}
    \def\title{twitter}
     \SAMPLEFIGURE
    \def\csvfilename{twitter_mpi-estime.csv}
    \def\title{twitter-mpi}
     \SAMPLEFIGURE
    \caption{The scalability of \emph{AC3Trim, AC4Trim} and \emph{AC6Trim} by using 16 workers. The number of edges is varied by randomly sampling from 10\% to 100\%}
    \label{fig:sample-edge}
\end{figure*}

Analogously, we show the result of sampling vertices in Figure \ref{fig:sample-ver}, in which the x-axis of plots is the ratio of sampling vertices. There are almost the same trends as shown in Figure \ref{fig:sample-edge}. One difference is in upper three plots; that is, over \emph{twitter} and \emph{twitter} we can see \emph{AC4Trim} traverse more edges than \emph{AC6Trim} as the the unsampled vertices are set to $\DEAD$ and their out-degree counters are still calculated.
The other difference is in lower three plots; that is, over \emph{twitter} and \emph{twitter} we can see \emph{AC6Trim} performs much better than \emph{AC4Trim}, except when vertices are 100\% sampled.
In this experiment, for implicit graphs loaded into memory, we can see that \emph{AC6Trim} is most scalable no matter how many vertices are trimmed and how many edges or vertices are sampled.   

\def\fwide{0.3}
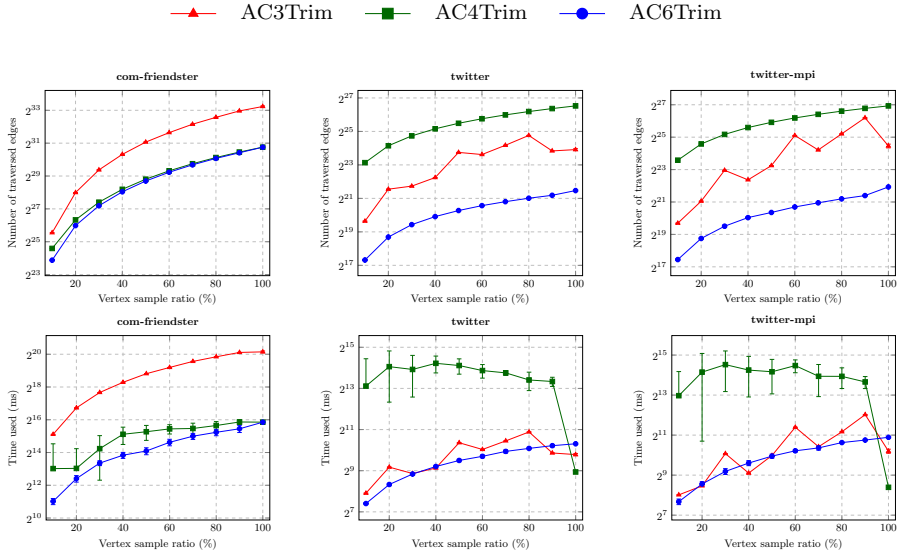
\begin{figure*}[!htb]
    \centering
    \begin{subfigure}[b]{1\textwidth}
    \centering
    \footnotesize
    \begin{tikzpicture}
        \begin{customlegend}[legend columns=5,legend style={align=center,draw=none,column sep=2ex},
                legend entries={{AC3Trim} ,
                                {AC4Trim} ,
                                {AC6Trim} ,
                                }]
            \addlegendimage{color=red, mark=triangle*,solid}
            \addlegendimage{color=darkgreen, mark=square*,solid}
            \addlegendimage{color=blue, mark=oplus*, solid}
           
        \end{customlegend}
    \end{tikzpicture}
    \end{subfigure}
    \def\yminmax{}

    \def\myxlabel{Vertex sample ratio (\%)}
     \def\myylabel{Number of traversed edges}
    \def\csvfilename{com-friendster-vsnum.csv}
    \def\title{com-friendster}
    \SAMPLEFIGURENUM
    \def\csvfilename{twitter-vsnum.csv}
    \def\title{twitter}
     \SAMPLEFIGURENUM
    \def\csvfilename{twitter_mpi-vsnum.csv}
    \def\title{twitter-mpi}
     \SAMPLEFIGURENUM
    
     \def\myylabel{Time used (ms)}
    \def\csvfilename{com-friendster-vstime.csv}
    \def\title{com-friendster}
    \SAMPLEFIGURE
    \def\csvfilename{twitter-vstime.csv}
    \def\title{twitter}
     \SAMPLEFIGURE
    \def\csvfilename{twitter_mpi-vstime.csv}
    \def\title{twitter-mpi}
     \SAMPLEFIGURE

    \caption{The scalability of \emph{AC3Trim, AC4Trim} and \emph{AC6Trim} by using 16 workers. The vertices are varied by randomly sampling at radio from 10\% to 100\%}
        
    \label{fig:sample-ver}
\end{figure*}

